# ELECTROPORATION OF BIOLOGICAL CELLS EMBEDDED IN A POLYCARBONATE FILTER

William A. Hercules, James Lindesay, Anna Coble
Computational Physics Laboratory
Howard University, Washington, D.C. 20059

Robert Schmukler
Pore Squared Bioengineering, Inc.
13905 Vista Drive, Rockville, MD. 20853

**ABSTRACT**

The electropermeabilization of biological cell membranes by the application of an external field occurs whenever an applied field exceeds a threshold value. For fields above this threshold value but less than another critical value, the pores formed in the membrane are transient or reversible. Several mechanisms have been proposed for the formation of these transient pores. Here we examine the local electric fields generated for the configuration of cells embedded in a polycarbonate filter, both in the region in and around the pore. We consider the shear forces created in the membrane due to the gradient of the field along the surface of the membrane, and the interaction of the charged molecules in the membrane with this field. A relationship between the electric field strength and the size of the pore formed is derived.

# Chapter I Introduction
## I.1 Membrane structure

All living cells have a membrane separating the cell contents from the cells environment. This membrane allows the cell to maintain an internal environment that is suitable for the survival of the cell. Since cells exists in some external environment, it is necessary that some form of interaction occur between the cell and its external environment. The cell membrane is a site of a large variety of cellular processes, ranging from communication, transport and excitability to intercellular interaction, morphological differentiation, and fusion.

The cell membrane functions as a barrier to the unrestricted movement of substances between the intracellular region and the environment of the cell. This barrier is selectively permeable to certain materials in the fluids surrounding the cells, allowing these materials to cross the membrane relatively unhindered while preventing the movement of others. It is thought that the selective permeability of the cell membrane for certain types of material in the intracellular or extracellular fluids is due to the presence of very small ionic channels which traverse the membrane and allow for the movement of small ionic species across the membrane. The cell membrane is composed of phospholipids arranged in an architecture that is commonly referred to as a lipid bilayer. Figure 1 shows a schematic of the cross section of the membrane illustrating the arrangement of the phospholipid molecules. The lipids present in the membrane are amphipathic, containing a polar head group and a non-polar lipid portion that is typically composed of relatively straight chain alkyl groups ranging from 17- 25 carbon atoms long. In this bilayer structure the lipid portions are apposed leaving the charged regions apart. This arrangement of the phospholipids results in an interior region of the membrane which is much less polar than the interfacial region.

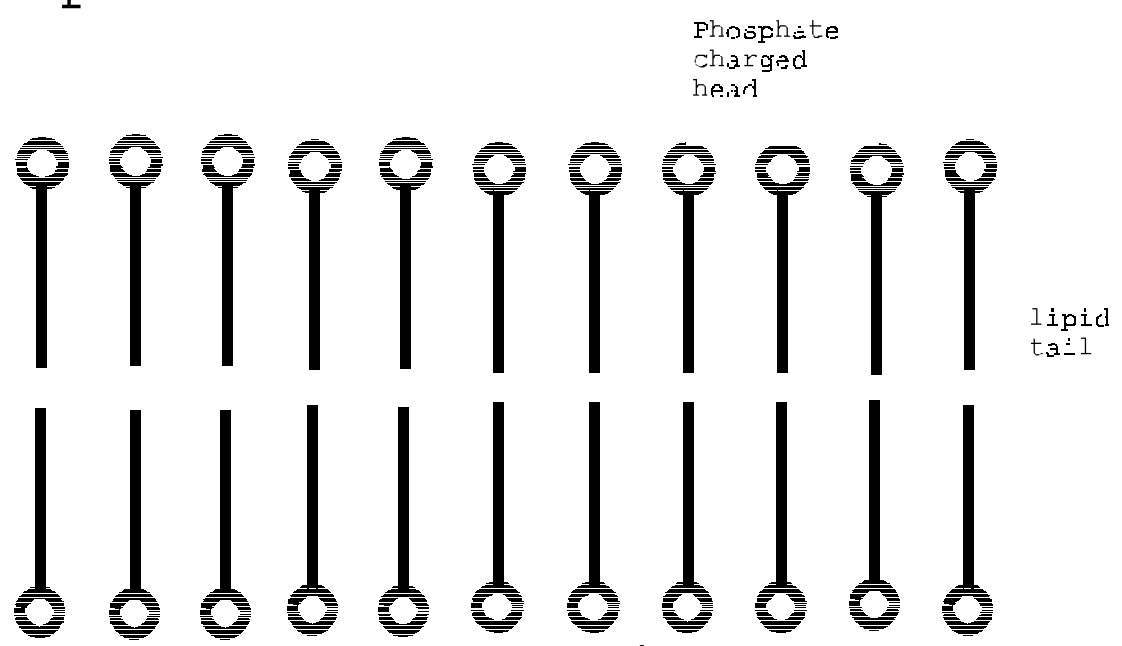

Figure 1. Schematic of membrane structure

Isolated membranes are white flocculent solids of density 1.05 - 1.35 $gm/cm^3$ [40]. Membranes are relatively unstable structures, not only in the sense that they fuse and that their components undergo exchange, but also that they can be readily disrupted and dissociated [40]. They can be disrupted by sonication, freeze thawing, osmotic shock and extensive mechanical shaking. The membrane may be dissociated into its chemical components by treatment with detergents, phospholipases, mixtures of organic solvents and chaotropic agents [40].

The membrane components may be resolved into two main classes of compounds, lipids and protein. In addition to these two main classes, there is a third class of compound , carbohydrates, accounting for up to 10 % by weight of the membrane components. The membranes of different organelles and cell types contain on a dry weight basis 20-60 % protein, 30-80 % lipid and up to 10 % carbohydrates [40]. The components of the membrane matrix are held together largely by non-covalent forces. These phospolipids are arranged with their polar phosphate head groups aligned in the ionic medium of the interior and exterior of the cell, while the nonpolar lipid chain are apposed on the interior region of the membrane. The acyl chains of the phospholipids interact with each other and with proteins primarily by Van der Waals interactions, while the interactions of the polar groups at the interface are expected to be largely of the Coulombic and hydrogen-bonding types [40].

In the membrane, specific interactions between the components lead to a selective ordering and segregation of these components in the plane of the membrane. A consequence of this segregation of components may be the establishment of a long-range order and cooperativity within segregated domains. Separating these domains are regions of discontinuity and mismatch, referred to as patch boundaries. Due to the fluid-like state of the membrane, the molecules within the membrane matrix can undergo a variety of motions; rotational motion along the axis perpendicular to the interface, trans-gauche conformational changes in the acyl chains and lateral diffusion in the plane of the membrane. The geometric constraints are such that the polymethylene chain forms a hydrophobic cylinder. To maximize the interaction between neighboring atoms, a chain can assume a stretched all-trans conformation. This arrangement gives a cylindrical structure with a theoretical cross sectional area of about 18 $Å^2$ and yields an average increase in chain length of 1.27Å per methylene residue.

Human erythrocytes are relatively simple cells comprising nothing more than a plasma membrane, hemoglobin and electrolytes. The cells are shaped like bi-concave discs, with a diameter of approximately 7 μm. The mass ratio of protein to lipid in the membrane is about 1.2:1 with the lipid component made up of glycerophospholipids, sphingomyelin and cholesterol in a mass ratio of about 1:0.3:0.5. There are two major intrinsic membrane proteins, both of which traverse the bilayer. In addition to the intrinsic protein, a network of extrinsic protein are associated with the cytoplasmic surface of the membrane. The membrane also has binding sites for numerous hormones, various enzymes for active ion transport, and mechanisms to facilitate the passage of small molecules such as glucose and glycerol.

The membrane skeleton comprises two major filamentous proteins, spectrin and actin, and a number of globular proteins. In the erythrocyte membrane the actin firms short filaments which link spectrin tetramers together to form a continuous network. The majority of the spectrin dimers are attached to the membrane by means of a protein known as ankyrin, a large globular protein which links the flexible end of spectrin proteins.

## I.2 Electroporation

Electroporation is the phenomenon in which a cell exposed to an electric field is permeabilized, as if aqueous pores are introduced in the cell membrane. The usual method of performing electroporation is to place cells in a suspension medium between the plates of a parallel plate capacitor and apply a potential difference between the plates of the capacitor.

Stampfli observed the breakdown of the membrane of nodes of Ranvier by the application of an electric field [78]. Sale and Hamilton observed that treatment of cells with intense electric pulses led to cell lysis [65] and suggested that the membranes of the cells were damaged by the transmembrane potential induced by the applied electric field. Further studies by Neumann and Rosencheck [55] ; Benz, Zimmermann et. al [9], Kinosita and Tsong [42] showed that the cell membranes of pulse treated cells were permeable to molecules with a size less than a certain limit, which suggested that some porous structure was created in the membrane. While the structure of the pores created by electroporation is not known, it is generally believed that they may be transient aqueous pathways that perforate the membrane. While most of these transient pores are believed to close rapidly, some may remain open orders of magnitude longer than the duration of the pulse. Diffusion and movement directly due to the electric field ( eg. Electrical drift and electro-osmosis ) are likely mechanisms of molecular transport across the cell membrane due to electroporation [56]. While the mechanism of electroporation is not completely understood, numerous experiments show that electroporation occurs for short pulses when the transmembrane voltage reaches approximately 1 V. The electric field pulses causing the electroporation of cells are typically of magnitude 1-20 kV/cm and have duration of 10 μs to 10 ms. Visual demonstrations of the pores formed in the cell membranes were not made until the electron microscope studies of Chang and Reese [14] in 1990, due to the difficulty of detecting the pores, which are small, transient structures.

A few studies have made quantitative measurements of molecular transport ( Mir et al., [48] 1988; Chakabarti et al., [12] 1989; Lambert et al., [43] (1990 ) particularly on a transport per cell basis ( Bartoletti et al., [5] 1989 ). Quantitative study of molecular transport is necessary in order to better understand electroporation. These studies are needed to ;

- suggest and test theoretical models
- provide a basis for comparing data from different molecules and experimental conditions
- guide applications of electroporation in research, biotechnology and medicine.

Most electroporation studies have used one of the following four methods of analysis ;

- expression of an introduced gene
- total population methods
- image analysis
- flow cytometry

Kinosita et. al.[43,44] used voltage sensitive dyes to measure the transmembrane potential required to cause poration. In their analysis they calculated the transmembrane potential that existed when the cell was placed in a uniform external electric field. The primary event that leads to pore formation is believed to be the induction of a large transmembrane potential by the applied electric field ( Kinosita et. al. 1992 ). For a spherical cell of radius, $a$ , in a uniform electric field, $E_0$ , the potential difference, $\Delta V$ , between the intracellular and extracellular surfaces of the cell membrane (in the absence of pores) is assumed to be given by

$$\Delta V = \frac{3}{2} f a E_0 \cos\theta \left[ 1 - \exp\left( -\frac{t}{\tau} \right) \right] \quad (1.1)$$

with

$$\tau = a f C_m \left( r_i + \frac{r_e}{2} \right)$$

$$f = \frac{1}{\left[ 1 + a G_m \left( r_i + r_e \right) \right]}$$

where, t, is the time after the constant field is turned on, $C_m$ is the membrane capacitance per unit area, $r_i$ and $r_e$ are specific resistances of the intra and extra cellular media, and $G_m$ is the membrane conductance per unit area [39]. Following the application of the electric field, the potential undergoes an exponential rise according to the equation above, and the steady state transmembrane potential difference reached from the equation above is the value given by the steady state solution to Laplace's equation (where $G_m \to 0$ so that $f \sim 1$).

$$\Delta V = \frac{3}{2} E_0 a \cos\theta \quad (1.2)$$

When biological cells are placed in an external electric field, the transmembrane potential, $\Delta V$, can be visualized using the voltage sensitive fluorescent dye, stryl dye RH292 following Grinvald et. al. [30]. When added to the extracellular medium, the dye (which has a hydrophilic head and two hydrophobic tails) is expected to orient itself with its head in the extracellular fluid. When the dye is bound to the membrane, it fluoresces brightly with an intensity of the fluorescence that is sensitive to $\Delta V$. The image of the fluorescence therefore indicates the spatial variation of $\Delta V$. Using a pulsed laser fluorescence microscope, the cells were imaged at a temporal resolution of 0.3 µs. The resulting membrane potential was found to follow the equation (1.1) for the expected spatial distribution of the potential.

For low electric field strengths ~ 100 V/cm the steady state potential in the membrane was achieved in a time of less than 2µs from the application of the external field. It was also found that the membrane could withstand a trans-membrane potential of up to 0.75 V, for at least tens of microseconds. When a larger external field was applied, for which the calculated $\Delta V$ was greater than 1 V, the observed behavior of the fluorescence was different. The actual change observed was much smaller than calculated and indicated saturation of the membrane potential. The spatial distribution of the potential deviated from the theoretical cos $\theta$ dependence and shows a flattened top and bottom indicating saturation of the membrane potential. The saturation level was found to occur at a potential of ~ 1 V. After a period of ~ 2µs the fluorescence intensity level slowly returned toward the original level on a microsecond time scale. The interpretation of these results was that when $\Delta V$ reaches a critical value of about 1V the membrane is porated and starts to conduct current. The conduction prevents any further increase in the membrane potential. The slow return of the fluorescence is interpreted as the gradual increase in the number of pores and/or the pore size. In RBC's the poration takes place within 1ms after $\Delta$V reaches a critical value of about 1V. Longer pulses are thought to produce larger pores or a greater number of pores. When $\Delta V$ exceeds 1V, a large current flows across the membrane because of the much larger conduction of the pore compared to the membrane.

Coster and Zimmerman [21] (1975) directly injected current into a cell of *Valonia Utricularis* and measured the transmembrane potential. When the potential, augmented by the injected current reached a critical value of about 1V, any further increase was counteracted by a dramatic increase in the membrane conductance, which was just large enough to keep the potential at the critical value. When the current was held constant after the breakdown , the potential gradually decreased on a microsecond time scale. The interpretation of this was that once the trans-membrane potential reached the critical value, the membrane became locally permeable. This increased permeability results in a large increase in the conduction across the membrane, which was enough to prevent the further increase in the transmembrane potential.

Experimentally, electroporation caused by short pulses universally occurs at a transmembrane voltage of about 1000 mV for many different types of cells and artificial membranes. Initial attempts to explain the rupture of plasma membranes were based on a theory involving the normal oscillating compression of the entire membrane and the resulting instability produced by the electrocompression of specific regions. Chizmadzhev et. al. suggested that pores were caused by local defects in the membrane. Similar pore approaches were developed independently by Weaver and Mintzer '81 [], and Sugar '81[80]. The models developed gave reasonable predictions for a critical transmembrane voltage for rupture, $V_{c,}$ but none could simultaneously describe the phenomenon of rupture and reversible electrical breakdown in charge injection experiments. Reversible electrical breakdown is not a true dielectric breakdown since there is not enough energy to ionize molecules. Instead it can be understood as a rapid electrical discharge due to the high ionic conductance caused by a gentle structural membrane rearrangement, that is pore formation. More recent theories provide a more reasonable explanation of the pore formation process. The general elements of these models are,

i) a resting membrane conductance containing an equilibrium population of hydrophilic pores,
ii) a membrane conductance G(t) due to the pores,
iii) transmembrane voltage V(t) induced changes in the pore population,
iv) The transient behavior determined by calculating how the pore population changes with V(t),
v) feedback between V(t) and G(t) involving both the pore population and external electrical resistances of the experimental system,
vi) Reversible electrical breakdown ( REB) caused by large ionic conduction through the transient pore population,
vii) Rupture of the plasma membrane caused by the appearance of one or more large unstable pores.

Hibino et. al.[34] suggests that if the electroporation uniformly increased the membrane conductance over the entire cell surface, then $\Delta V$, in the porated cell should still be proportional to cos $\theta$, where $\theta$ is the angle between the direction of the applied field and the position on the cell. Their experimental data, however, showed that was not the case. The finite membrane conductance caused by the pores was introduced in the two regions facing the electrodes of the external applied field. The spatial distribution of the conductance could be described by the function $G(\theta)$;

$$G(\theta) = \frac{G_0\left(|\cos\theta| - \cos\theta_c\right)}{1 - \cos\theta_c} \qquad 0<\theta<\theta_c,\ \ 180-\theta_c<\theta<180$$

$$= 0 \qquad \theta_c<\theta<180-\theta_c$$

where,
$G_0$ is the maximal conductance of the membrane and $\theta_c$ is the angle at which the transmembrane potential reaches the critical value.

This equation suggests that the conductance of the membrane only increases in the region of the membrane within an angle of $\theta_c$ of the direction of the applied electric field. Implicit in this equation is the suggestion that only those regions of the cell membrane where the potential exceeds the critical value, $\Delta V_c$, are porated, and that the conductance of different regions of the membrane are symmetrical around the direction of the applied electric field. It means that the conductance introduced is proportional to the excess potential that could be attained in the absence of pores minus the critical potential for poration. $\theta_c$ is interpreted to be the angle at which $\Delta V$ reaches the critical value for poration. The value of $G_0$ depends on the intensity and duration of the applied field. For field intensities of a few hundred V/cm and field durations of tens of microseconds, $G_0$ was in the range 1-10 $S/cm^2$ both at high and low ionic strengths. This observed conductance is calculated to be equivalent to the replacement of approximately 0.1 % of the total membrane area by aqueous openings. The above equation implies that the two regions opposing the electrodes are equally conductive. This is confirmed by the experimental indication of $\Delta V$ in fluorescence experiments that both regions are equally fluorescent. This can be explained from a theoretical view as follows. If one side of the cell is porated, immediately the potential difference across the other end of the cell increases to the critical value. This results in the other side of the membrane becoming porated. This means that the conductance of the membrane is also symmetric, the symmetric G does not mean that the changes taking place in the membrane are the same for each side of the cell, since different combinations of pore size and number of pores can produce the same conductance. Kinosita et. al. found evidence to indicate that the structural changes taking place on each side of the cell was observed to be asymmetric.

The poration takes place well within 2µs of the transmembrane potential reaching the critical value. Kinosita and Tsong [42] (1977) found that in red blood cells (RBCs), the poration takes place within 1 µs after the transmembrane potential reaches the critical value of 1 V. In a low salt , low ionic strength, extracellular medium, where the rise of the potential is relatively slow, the poration takes place at the moment that the transmembrane potential reaches the critical value of approximately 1 V. This critical potential of about 1 V for poration applies only to short pulses. Kinosita et. al. found that an applied field of 100 V/cm, which gave a theoretical transmembrane potential of 0.75 V , applied for 1 ms produced the same saturation of the membrane potential as did higher fields applied for a shorter duration. This suggests that the combination of the time and the intensity of the applied field determines the poration in the membrane.

After poration, the recovery of the cell membrane (in the case of reversibly porated membranes) can be inferred from measurements of the transmembrane potential. Using pulses of magnitude less than that required to porate the membrane, the response of the membrane to these pulses was examined. If the response of the membrane to the subcritical probe pulse was the same as the response of an intact cell, the porated cell was determined to have recovered, since it now showed no signs of increased conductivity, or pores in the membrane. Using this method, Kinosita et. al. determined that for sea urchin eggs treated with an electric field of 186 V/cm, 20 µs pulse, the cells recovered within 2 s [43] ( Kinosita 1988 ). However, immediately after poration the membrane was still highly conductive. At a time of 20 µs after the pulse, the conductance of the membrane, $G_0$ measured with a probe pulse of 67 V/cm was determined to be as high as 0.3 $S/cm^2$. This value may well be due to the conductance of the pores in the membrane, as well as partially closed pores reopened by the test pulse. After a time of 1 ms the value of $G_0$ decreased by an order of magnitude.

In studying the influx of fluorescent dyes into porated eggs after exposure of the egg to a long intense pulse ( 800 V/cm for 1ms ), it was found that relatively large molecules such as RH292 or its zwitterionic relative RH160 were seen to penetrate into the cell interior. Penetration was slow , requiring seconds in low salt medium, to minutes in $Ca^{2+}$ free seawater.

Although the field induced potential, $\Delta V$ is symmetric with respect to the direction of the applied field, it has often been found that the permeation of ions or molecules through the porated membrane have been asymmetric (Rossignol et. al. 1983 [65]). In sea urchin eggs, the dye and $Ca^{2+}$ permeation was much higher on the negative-electrode side, whereas the dye burst was seen mainly on the positive side of the liposomes. It was also found, that in liposomes in particular, the formation of holes in the membrane in the presence of the field was always symmetric.

To understand the phenomenon of electroporation many questions must be addressed.

- How does an electric field interact with the cell membrane?
- What are the subsequent events leading to the breakdown of the membrane permeation barrier?
- What would happen to a cell during the time that the

membrane function is disrupted?

- How would a cell then repair the damaged membrane and restore its normal functions?

Tsong states that the field interaction with the membrane is electrostatic, since any other interaction would violate thermodynamics. The potential differences required for electroporation of the membrane ( ~ 1V ) is not sufficient to cause true dielectric breakdown of the membrane.

Gross, Loew and Webb [31] (1986), used changes in the fluorescence intensity of a charge shift potentiometric dye incorporated into the cell plasma membrane and measured by digitally intensified video microscopy to measure the changes in the cell membrane potential induced by the application of an external electric field. Optical methods used for the measurement of fluorescence or absorption was first reported in the early 1970s. In their study Gross et. al. used a new stryl fluorescent indicator of membrane electric field 1-(3-sulfonatopropyl)-4-[ß] [2-(di-n-butylamino)-6-napthyl vinyl] pyridinium betaine, (di-4-ANNEPPS) developed by Loew and colleagues. They used the probe because it responded linearly to membrane potential at a level of ~ 9 % fluorescence change per 100 mV change in membrane potential. It also had characteristics that made it suitable for microscopy, since it could also be incorporated reversibly into one leaflet of an artificial bilayer. The molecule also possessed no net charge at neutral pH and the mechanism of its response to membrane potential was thought to be electrochromic in nature.

Artificial membranes provide a method of studying the properties of the cell membrane. A BLM formed from the lipid extract of cell membranes is a good permeation barrier for ions and hydrophilic molecules. The membrane specific conductance for $Na^+$ or $K^+$ is usually smaller than $10^{-8}$ S /$cm^2$. Since lipid molecules are either charged or polarizable, an electric field imposed across the membrane will disturb the highly ordered bilayer arrangement of lipids. For a BLM of a single lipid component, the conformation of the lipid molecules is controlled. At low temperatures, the molecules are arranged with $CH_2$- $CH_2$ bonds aligned in a trans configuration in the hydrocarbon chain. When the temperature is raised above a critical point, $T_c$, many of these bonds are converted to cis conformation. These cis bonds can cause the bilayer to become disorganized [82] (Thompson et. al. ). As a result, for temperatures above Tc, the bilayer exists in the fluid like phase ( F-state). A bilayer in the F-state has a higher conductivity for ions than when it is in its usual S-state. Experiments show that the permeability of a lipid bilayer to ions is maximal at Tc, a temperature for which the lipid exists half in the S-state and half in the F-state [83] (Tsong et. al . 1977), [26] (El-Mashak and Tsong) (1985).

Neumann et. al. [55] suggests that studies using electro-optic and conductometric measurements are consistent with electric field induced changes in the membrane structure of the vesicles. Relatively few electrical measurements have been made on isolated cells, because the measurement of the transmembrane voltage, $\Delta V$, requires the use of either microelectrodes, [9] (Benz and Zimmermann '79 ) or voltage sensitive dyes [43] ( Kinosita et. al. '88 , [46] Kinosita et. al. '92 ).

Studies of pore formation in artificial bilayer membranes have demonstrated four different types of behavior. These behaviors may be classified in terms of the effect on the membrane. For the largest pulse sizes the membranes showed irreversible electrical breakdown, in which the membrane discharged to a potential of zero. For pulse sizes slightly smaller, the membrane showed incomplete reversible electrical breakdown, in which the membrane discharged to a potential different from zero. For still smaller pulses, the membrane showed mechanical rupture, and showed a slow, sigmoidal shaped electrical discharge. For the smallest pulse sizes, the membrane was charged without any dramatic change of the membrane potential. The ability of a single membrane to exhibit four different outcomes by changing only the electric pulse characteristics is striking.

Initial attempts to explain the rupture of planar membranes were based on electrocompression [22 ] ( Crowley '73). This approach could not explain the rupture of "solvent free" and biological membranes, which have small values of compressibility. More complex models based on the viscoelastic behavior of a membrane, identified reversible permeabilization or REB, with the instability of the membrane, even though the subsequent behavior of the membrane potential was not treated. A theoretical model for pore formation in artificial membranes was proposed by Sowers and Lieber [77] utilizing a statistical model. Their model was based on the normal thermodynamic distribution of energies in the molecules with those regions having above average energies being more likely to spontaneously form pores.

## I.3 Theories of Pore Formation

Crowley [22] ('73) first examined the effects of a high, transverse, electric field on the mechanical equilibrium of a flat isotropic, elastic membrane sheet. Using linear stability analysis, he demonstrated that when the interfaces of the membrane are perturbed by disturbance waves, the transverse electric field gives rise to electric surface stresses which destabilize the system and cause squeezing deformations of the bilayer. Crowley obtained an expression for the compression of the membrane by equating the squeezing Maxwell tension to the elastic stress due to the transverse strain. Crowley proposed that pore formation results when the membrane potential exceeds the threshold value, and the growth of the squeezing disturbance eventually causes the membrane to rupture, locally creating a hole. His results gave the correct order of magnitude for the critical potential, $V_c$ , but underpredicted the critical potential by approximately a factor of 2.

His approach could not explain the rupture of "solvent free" and biological membranes, which have smaller values of compressibility. The explanation of electroporation as a linear, electromechanical instability has since been criticized in the literature, primarily because it predicted an inordinately large value for the compression of the bilayer at the critical potential. Experimental results and other lines of theoretical study suggest that the compression is no more than a few percent [54].

Maldarelli and Stebe [48] proposed a re-examination of the study of the theory of electropore formation based on the instability mechanism of the membrane due to the mechanical anisotropy of the lipid bilayer. They concluded that in the theory proposed by Crowley, the incorrect predictions for the unacceptably large electrocompression at the critical potential could be traced to an incorrect assumption that the lipid bilayer is isotropic in its elastic response.

Abidor, Chizmadzhev and co-workers suggested that local defects (pores) in the membrane are responsible for the reversible breakdown of the membrane [1].

One point that has been established about pore formation and cell fusion is that they are result of the field induced transmembrane potential. Under the influence of an external electric field, the potential drop across the cell is sustained by the cell membrane, which is a poor conductor, while the field in the much more conductive ionic medium of the interior of the cell is small.

Studies were carried out by Sowers and Lieber [77] on the size, lifetimes, numbers and location of the pores formed in erythrocyte membranes using low level video microscopy techniques on RBC ghosts. Two sets of experiments were carried out on labeled membranes. The first was to determine the effect of a 1.2 ms pulse, the second was to determine the effect of a 0.6 ms pulse. To probe the diameter of the electropores they used soluble fluorescent molecules. These included 6-deoxy-N-(7-nitrobenz-2-oxa-1,3-diazol-4-yl) aminoglucose ( NDB-G) , several fluorescent dyes of average molecular weight 10000, 70000, 156000 Daltons. These molecules have average unhydrated and unconjugated sphere radii of 0.45, 0.54, 0.75 ( from C-C bond lengths), 2.3, 5.8, 8.4 ( from chromatography ), and 1.8, 2.8, 4.3 from ( 0.00151 $nm^3$/Da ).

In their experiments, they found that regardless of the pulse characteristics, efflux occurred for all fluorescent labels except fluoresceinated bovine serum ( FB) , fluoresceinated myoglobin (FM) and R-phycoerythrin (RP), and that whether the pulses were 0.6 ms pulses or 1.2ms pulses, each pulse always induced only some of the fluorescence to leave the ghost. Additional pulses eventually caused the membrane to lose all visible fluorescence. This suggested that after induction the electropores were self resealing, i.e. reversible pores, at least to a radius equal to or smaller than the size of the probe molecule. It also indicated that the labels used did not bind to the membrane. They also found that a distinct cloud of fluorescence appeared outside the hemisphere facing the negative electrode after a single 1.2 ms pulse. The cloud developed 20-50 ms after the pulse and was visible for about 200 ms after the pulse. During the existence of the cloud they estimated that the total pore area as a fraction of the total membrane area was about 0.0006.

Since the largest dextran used ( m 156000 ), was able to escape the cell via the pores formed in the membrane, the pores were determined to be at least 8.4 nm in diameter, corresponding to the size of the largest dextran molecule. The loss of fluorescence occurs in increments; a large immediate loss over a very short period followed by little or no loss over a much longer interval. This suggests that the pore opens to some finite size for a finite time and then reseals rapidly. They observed continuous loss of fluorescence only when NDB-G was used as a probe, suggesting that the pores must reclose to a residual radius which is smaller than LY (0.75nm) but larger than NBD-G (0.45 nm) and remain at that radius. They concluded that the formation of electropores were unlikely to be involved in the electrofusion mechanism. Other studies measured the effective diameters and provide an estimate of the numbers of the electropores formed.

Bliss et.al [13] (‘88) used flow cytometry to measure properties of RBCs. Flow cytometry is useful for providing rapid optical measurements, both light scattering and fluorescence, on large numbers of individual cells. It also provides a means of investigating morphological changes and uptake of fluorescence-labeled test macromolecules caused by electroporation. The use of fluorescence measurement in flow cytometry provides a flexible and sensitive analytical technique in biological investigations. The forward blue scatter ( FBS) signal from the flow cytometry is widely used as an indicator of the size and shape of cells [51,75,84] The FBS signal with a threshold is used to establish whether a significant optical event had occurred.

In these experiments, a single 50 µs pulse was applied to the cells in a PBS suspension containing 0.1mM FITC-dextran (70,000 dalton). The test molecule was chosen because it has a relatively small electrical charge. The FITC -dextran provides a good probe for the existence of relatively large openings or pores in the cell membranes. The applied electric field was determined from measurements of the current flow, and using the medium resistivity and the geometry of the chamber. This overcame the possibility of large voltage drops which can occur between the electrode-electrolyte interface. Their results indicated that the percent of electroporated cells with large pores, as revealed by the significant uptake of FITC-dextran, was greater than 20% for even a single pulse of optimal amplitude ( 8.8 KV/cm ) .

Klenchin et.al [46] (1991), using transfection of Simian Cos-1 cells with DNA, concluded that the uptake of the DNA by the cells following the application of a porating pulse was electrophoretically driven. In their experiments they found that the uptake of the DNA was an order of magnitude larger when the electric field was oriented so that the DNA moved in the opposite direction of the field toward the cells as compared to when the direction of the field was such that the DNA had to move in the direction of the field in order to enter the cells. DNA is slightly negatively charged. Their results also suggests that the uptake of the DNA by the cell is a fairly rapid process that occurs within a time period of less than 3 seconds. If the DNA is placed in the bathing solution prior to the application of the pulse they found that the transfection rate was much larger than if the DNA was introduced after the application of the pulse. The efficiency of penetration of the DNA into the cell was found to be two orders of magnitude higher in the case where the DNA is introduced before the pulse as compared to when it is introduced after the pulse.

Their results indicated that there was a characteristic time for the penetration of the DNA into the cells. This time was estimated to be approximately 3 seconds following the application of the pulse. This suggests that the pores formed in the membrane are only able to allow the passage of the DNA for a time of 3 seconds. They also investigated the effect of varying the electrophoretic mobility of the DNA in the pulsation medium. Ficoll-400 was added to the pulsation medium to increase the viscosity of the medium and hence decreases the mobility of DNA. Other factors that may also affect the mobility of the DNA in solution include the effective charge on the molecule and the ionic environment. Results indicated that as the concentration of the Ficoll-400 in the medium was increased the transfection efficiency of the DNA decreased. One can conclude then that electrophoresis may play a significant role in the penetration of DNA into the cell. They also suggested that it may be possible for the DNA to enter the cell even when the size of the pore is less than the size of the DNA molecule. They proposed that the presence of the field may allow the DNA molecules in contact with the cell membrane to produce a force on the molecule that causes the membrane to create an evagination so that the DNA with the surrounding membrane sac is taken into the cell.

## Chapter II Experimental basis of present theory

The normal method for performing electroporation is to place the cells in suspension within a region of an externally applied electric field. The electric field is usually set up by the plates of a parallel plate capacitor, with plate separation of from approximately 0.1 to 1 cm. The cells undergoing electroporation are placed in a suspension that fills the spaces between the plates of the capacitor. In the capacitor space, the cells are freely mobile and can translate as well as rotate in the field. This freedom of movement of the cells means we have no specific information about the position or orientation of the cells when it is exposed to the porating electric field. This makes it difficult to determine the particular region of the cell exposed to the maximal electric field.

The transmembrane potential $\Delta V$ for a spherical cell in a uniform field it has been shown in eq. (1.1). For a typical cell of 1 μm radius, to achieve a transmembrane potential of ~ 1V, the applied field E, should have a magnitude of approximately $1 \times 10^6$ V/m or $10^4$ V/cm. In order to obtain the required magnitude of the field with a plate separation of from 0.1 to 1 cm, a voltage between the plates of $10^3$ to $10^4$ volts is required. This magnitude of a voltage results in large currents and Joule heating effects when the suspending solution is normal saline solution. The Joule heating can result in severe damage to the cells in the suspension medium.

In one approach used to minimize the effect of the Joule heating, the cells are placed in a suspension medium which is much less conductive than normal isotonic saline solution. This approach leads to other problems, in that the cells are now no longer in their normal environment, and are thus likely to suffer damage to the cell membrane from exposure of the cell to a foreign environment. While using less conductive solution may address one problem, it does not address the free motion of the cells in the suspension, so that their orientation in the field is still unknown, and thus attempts to relate the position of pores formed in the membrane, to the position of the maximal transmembrane potential is difficult.

The electroporation method we utilize differs from the normal method of performing electroporation. This technique makes use of a polycarbonate chamber in which is placed a very thin ( ~ 10 - 15 μm ) polycarbonate filter [ 71]. The system consists of a cylindrical polycarbonate chamber formed from mirror halves that can be separated at the middle. Figure 2 shows a vertical sectional view of the chamber. The chamber has an inner lumen of 1 centimeter diameter and a height of 2 cm. A thin polycarbonate filter can be placed in the middle of the chamber.  The filter serves as a partial barrier between the upper and lower parts of the chamber since it has small cylindrical pores which extend through its thickness,.

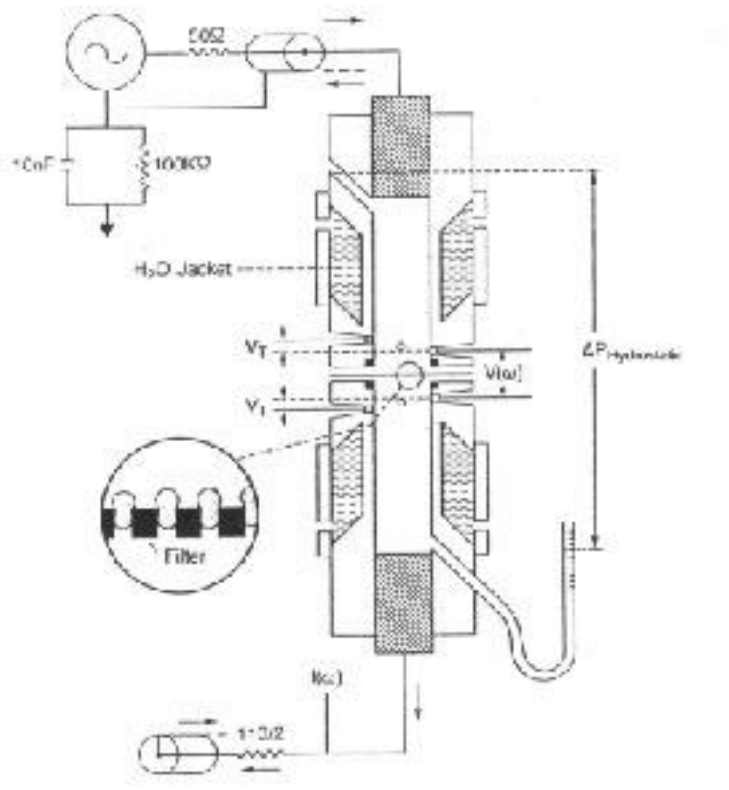

Figure 2 Schematic of Impedance Chamber. (From: Schmukler [70] )

Located at the top and bottom of the chamber are carbon/graphite electrodes which are used to apply an electric field to the contents of the chamber. The side walls of the chamber have openings for the insertion of voltage measuring electrodes. These measuring electrodes are positioned on either side of the filter so that the potential difference across the filter can be measured.  The top and bottom parts of the chamber have ports that permit the chamber to be individually filled with isotonic saline solution.  These also provide ports for the introduction of other materials into the chamber.  The filter is placed in the middle of the chamber which is then filled with isotonic saline solution.  When a potential is applied across the ends of the chamber, a substantial potential difference forms across the filter. Since the filter is less conductive than the surrounding saline solution, the potential difference across the filter is a significant part of the applied potential.

When cells are trapped in the pores of the filter, the portion of the cell membrane that is embedded in the filter pore is exposed to approximately the same potential difference that exists across the filer. This potential difference may be reduced because of the current leakage around the cells trapped in the pores. If however the leakage is relatively small, then the cell membrane that is positioned in the pores of the filter experiences the potential difference across the filter.  This potential difference across the thin filter (~ 15 μm) results in a large field across the filter.

The present technique overcomes some of the problems of electroporation in suspension, while introducing others. One of the main advantages of the present method is that the cells are held In a fixed position by embedding them in the pores. Because the cells are held in a fixed position relative to the external field, it is possible to make determinations of the location of pores formed in the membrane relative to the direction of the applied electric field. Another advantage is that the external electric potential is small enough to minimize Joule heating.

Since the cells are also placed in an isotonic solution the surrounding environment is similar to the environment that the cell normally experiences. In addition, since lower potentials can be applied across the chamber to achieve the same magnitude field across the cells embedded in the filter as that achieved in suspension media, it is possible to perform more precise measurements and obtain information about the response of the cell membrane than is possible using a suspension system.

With the given dimensions of the chamber, the field that exists between the electrodes is approximately uniform. In this procedure cells are allowed to sediment and become layered onto the filter and then made to embed in the filter by the application of a slight negative hydrostatic pressure ( ~ 40 cm $H_2O$). The application of the negative pressure causes a small portion of the cell membrane to be drawn into the pore of the filter. With the cells embedded in the filter, the electric field is applied simultaneously to the entire sample of cells ( ~ $3.3 \times 10^5$ cells ).

The protocol for an experiment with Red Blood Cells is as follows:

- The chamber is rinsed out with saline and thoroughly dried
- The filter is placed in the chamber and the chamber reassembled
- The chamber is carefully filled with saline solution to avoid stretching the filter and to avoid forming any air bubbles in the chamber.
- The Red Blood cells (RBC) are prepared from 5-6 drops of blood suspended in approximately 15 cc of saline solution. The mixture is shaken to ensure uniform mixing and then centrifuged at low speed for ~ 5 minutes to wash the RBC's.
- The supernatant is removed along with the top layer of cells, which contain the buffy coat. This layer contains most of the white blood cells.
- The cell precipitate is then re-suspended in 0.5cc saline and mixed thoroughly.
- Approximately 0.1 cc of the re-suspended cells are injected into the chamber just above the filter.
- The cells are injected into the chamber with the chamber upright and allowed to sediment onto, and

disperse across the filter, for approximately 5 minutes.

- The lower portion of the chamber is attached to a pressure head via a 3- way stop-cock.
- With the upper portion of the chamber open a downward positive pressure equal to approximately 40 cm of water is applied from the lower portion of the chamber by maintaining a lower hydrostatic level, in the connected pressure head.
- The applied pressure causes the cells to seat into the pores of the filter. The extent of the cell-filter seal can be determined by checking the flow rate of fluid through the filter. If the cells form a proper seal the flow rate of fluid across the filter is essentially almost zero.
- With the pressure on, the chamber is then inverted and the loose cells allowed to sediment away from the top of the filter. These loose cells are then washed away with saline.
- It is important to remember that the orientation of the chamber is inverted from the original position, so that the top of the chamber is now at the bottom

Using our method , the cells are held in a fixed position relative to the external applied field since they are embedded in the filter. The pores in the filter are made by etching the tracks left by alpha particles incident on the polycarbonate filter material to produce roughly cylindrical pores with an average diameter of 1.85 µm. The pores may have a maximum deviation in the axis of symmetry from the vertical of about 30°.

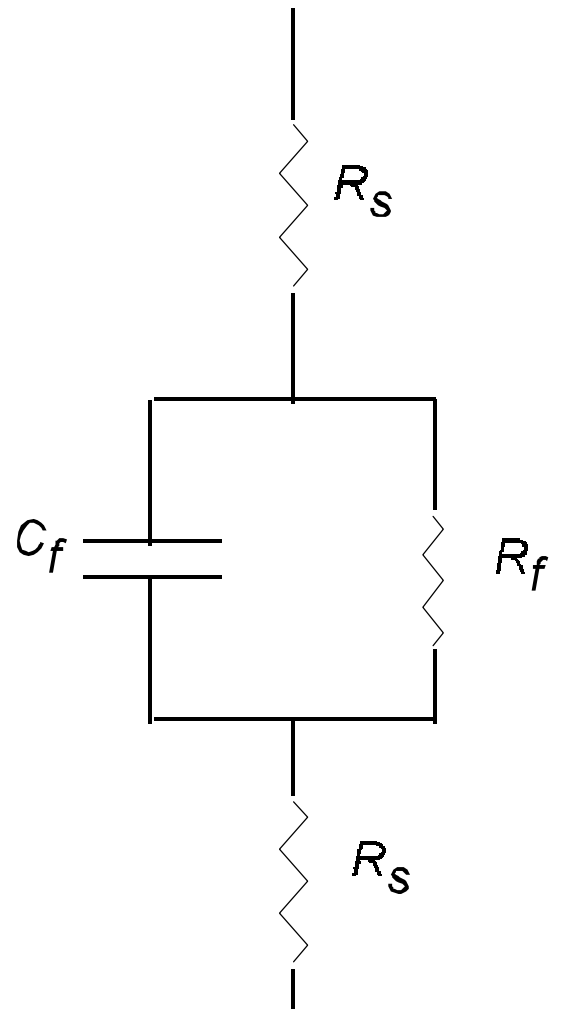

Figure 3a. Electrical model of the Chamber with filter.

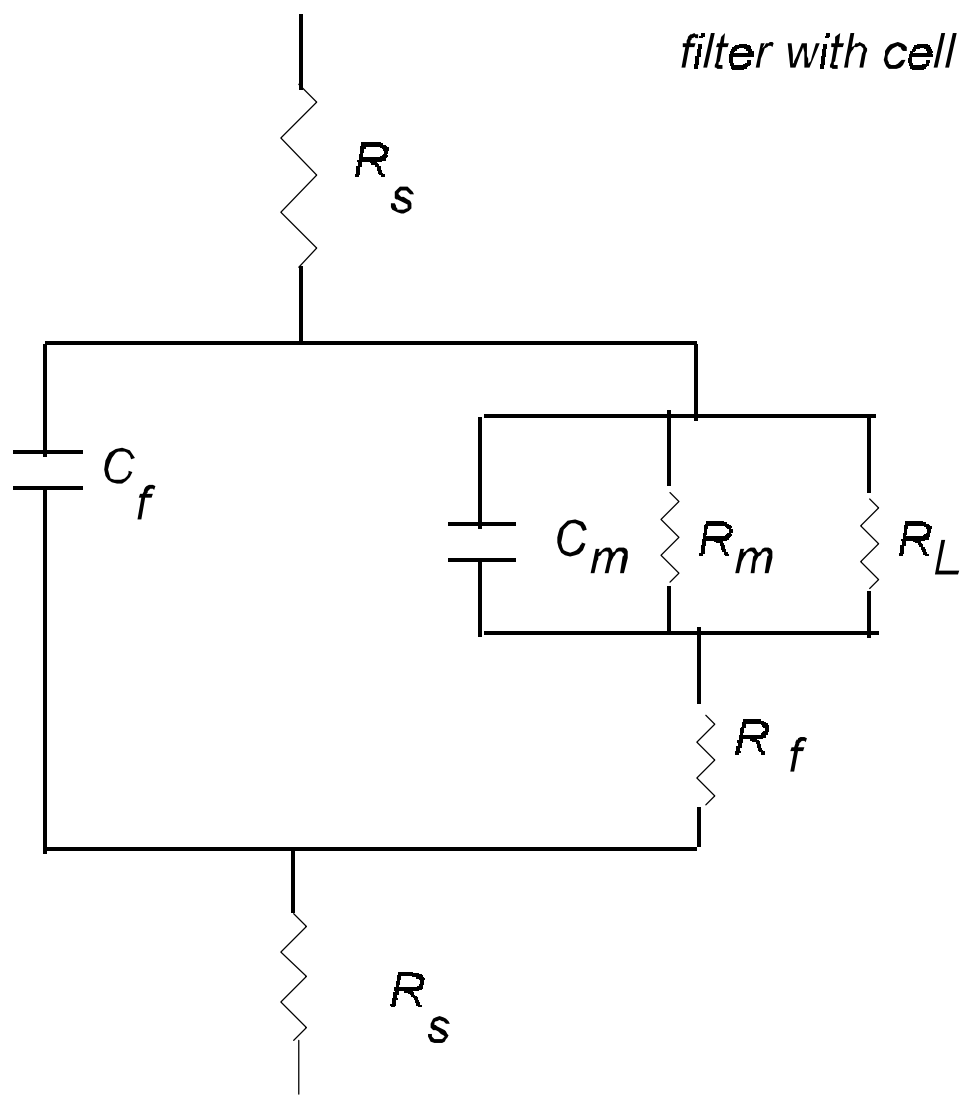

Figure 3b. Electrical model of chamber with filter and cells.
$R_s$=Resistance of soln. $R_L$=Leak resistance
$R_f$=Resistance of filter $R_m$=Resistance of membrane
$C_f$=Capacitance of filter $C_m$=Capacitance of membrane

The setup of the chamber with the filter in place may be modeled in terms of combinations of resistors and capacitors. The main resistances are due to the solution filling the chamber, the filter and the cell membrane. For the chamber setup without the filter, the system may be viewed as a resistance due to the solution. With the filter in place, the system can be represented in terms of the combined series resistances of the saline solution and the filter. The resistance of the filter may also be viewed in terms of the conductance due to the channels in the filter. Since the filter is made of a dielectric material, it also has some capacitance. Thus the filter may be represented electrically as a resistance or conductance in parallel with a capacitance.

The total conductance of the filter is due to the combined conductance of the pores in the filter. Thus from a determination of the resistance of the filter and a knowledge of the average number of pores in the membrane it is possible to determine the resistance of a single pore. With the filter in place, the chamber may be modeled as shown in figure 3a.

The addition of the cells to the chamber and the embedding of the cells into the filter pores modifies resistance and capacitance of the system. In addition to the cell contributions, there is a leakage current due to the layer of liquid between the cell membrane and the walls of the filter pore. With cells embedded in the filter the system may be modeled as shown in figure 3b.

Using differential measurements, we can determine the contributions of the various elements to the impedance of the system. An impedance measurement on the chamber, without the filter in place, gives the magnitude of the saline impedance between the voltage measuring electrodes. The addition of the filter gives the impedance of the saline and the filter. The difference between these two measurements gives the impedance of the filter only. The addition of the embedded cells permits us to determine the contribution of the cell membranes and the leak resistances around the cells.

Once the electropores are formed in the cells, an impedance measurement gives the total impedance of the system with pores. From this measurement, and the value of the system before the electropore is formed, we can determine the change in the impedance due to the formation of the electropores. This change may be used to model the formation of the electropores.

We seek to relate the pore size determined from the impedance measurement to that determined from a theoretical consideration. The simplest approach is to approximate the cell as a spherical cell and treat the problem as that of a dielectric sphere in a uniform external electric field. The major problem is then one of determining the potential distribution in the region of the pore as well as over the remaining surface of the cell.

In the present approach the simplest model of the pore is one of a cylindrical opening in a flat membrane. This model leads to a solution of Laplace's’ equation in cylindrical coordinates. Unfortunately the boundary conditions at the top and bottom of the cylinder can not be uniquely specified. This limitation leads to an incomplete specification of the solution. In cylindrical coordinates the solution involves a sum of Bessel functions. Due to these difficulties in describing the boundaries and potentials everywhere on the cylinder wall, the expansion coefficients cannot be uniquely determined.

## Chapter III Theoretical approach to the problem

The goal of our theoretical approach is to relate the size of the pore formed in the cell membrane to the magnitude of the applied field. We seek to determine the form of the electric field in the region of the electropore formed in the membrane. In considering the fields existing in the pore, we take the membrane to be a single layer of uniform conductivity. The internal structures of the cell as well as other membrane components are also ignored. The microscopic details of the cell membrane, i.e., membrane proteins, microcalyx, microtubules, membrane glycoproteins and other structures, are not included in this model. The possible impact of the cell nucleus where applicable, endoplasmic reticulum, mitochondria etc., on the field distribution is also ignored for this treatment.

The intracellular region is viewed as essentially a uniform medium with a single conductivity $\sigma_3$. The conductivity of the intracellular contents is taken to be much larger than the conductivity of the membrane. This is in accordance with the composition of the cell membrane as composed mostly of non-conducting lipid components. The intracellular components on the other hand are mostly ionic in nature, which results in the interior of the cell having a much larger electrical conductance than the membrane. We can therefore model the cell as a spherical shell with an outer radius of R and shell thickness, t.

The initial approach used was to solve Laplace's equation in spherical coordinates and thus determine the potential over the entire shell. Next, we would determine how the potential distribution changed when a pore was formed in one region of the cell. The pore was treated as a cylindrical pore in a flat sheet, which leads to a description of the pore in cylindrical coordinates. The potential distribution in the pore was sought. The form of the solution involves the Bessel functions, but because of limitations in specifying the boundary conditions on the cylinder, the analytical form of the solution could not be found. In addition to the difficulty in describing the boundary conditions in the cylindrical pore, we see that the description of the pore as a cylinder in a flat sheet is not a biophysically realistic model. Because the membrane is essentially a fluid sheet, the edges required in the cylindrical model would not be present. Thus, we seek an alternative description of the pore. The description of the pore and surrounding membrane should be such that it more realistically models the membrane around the pore. The model should account for the relatively constant thickness of the membrane as well as the fluid nature of the membrane itself. The membrane model geometry was expressed in oblate spheroidal coordinates. In this view, when the pore is formed, the upper and lower portions of the membrane must join around the lumen of the pore. Thus we expect that around the pore, the membrane must be greatly curved. We can therefore use the curvature of the membrane to specify the particular coordinate surface chosen to represent the membrane.

Due to the limitations of determining the analytic form of the potential in the pore from a solution of Laplace's equation, an alternate method is sought for obtaining a solution. The method uses Maxwells equations and boundary conditions on the field, as well as the continuity condition to determine a functional form for the potential in the pore. With the potential in the pore determined we find the current flow in the system. Through the continuity condition, we relate the current in the different regions of the system and thus the fields existing in the different regions. This allows the determination of the relationship between the applied field and the field in the electropore.

In order to achieve the theoretical goal of determining a relationship between the applied field and the size of the pore formed, we utilize a simple interaction between the field and the membrane. This interaction examines the force due to the inhomogeneity of the field in the pore and the charges in the charged heads of the phospholipid components of the membrane. We argue that this force is counteracted by the surface tension of the membrane such that the electropore will expand until the forces due to the field is equal to that due to the surface tension effects on the hole.

## III.2 Oblate spheroidal coordinates

We will briefly review the transformations needed to calculate the form of the fields in and around the pore. A more detailed calculation of the transformations is given in the references (Hercules, et.al., 2002). In oblate spheroidal coordinates the transformation equations from Cartesian coordinates can be written as

$$x = a \cosh u \cos v \cos \varphi = a \left(1 + \eta^2\right)^{\frac{1}{2}} \left(1 - \zeta^2\right)^{\frac{1}{2}} \cos \varphi$$

$$y = a \cosh u \cos v \sin \varphi = a \left(1 + \eta^2\right)^{\frac{1}{2}} \left(1 - \zeta^2\right)^{\frac{1}{2}} \sin \varphi$$

$$z = a \sinh u \sin v = a\, \eta\, \zeta$$

We will also need the scale factors which relate infinitesimal coordinate changes to infinitesimal lengths:

$$h_1 = h_\eta = a\left[\frac{\left(\eta^2 + \zeta^2\right)}{\left(1 + \eta^2\right)}\right]^{\frac{1}{2}}$$

$$h_2 = h_\zeta = a\left[\frac{\left(\eta^2 + \zeta^2\right)}{\left(1 - \zeta^2\right)}\right]^{\frac{1}{2}}$$

$$h_3 = h_\varphi = a\left[\left(1 + \eta^2\right)\left(1 - \zeta^2\right)\right]^{\frac{1}{2}}$$

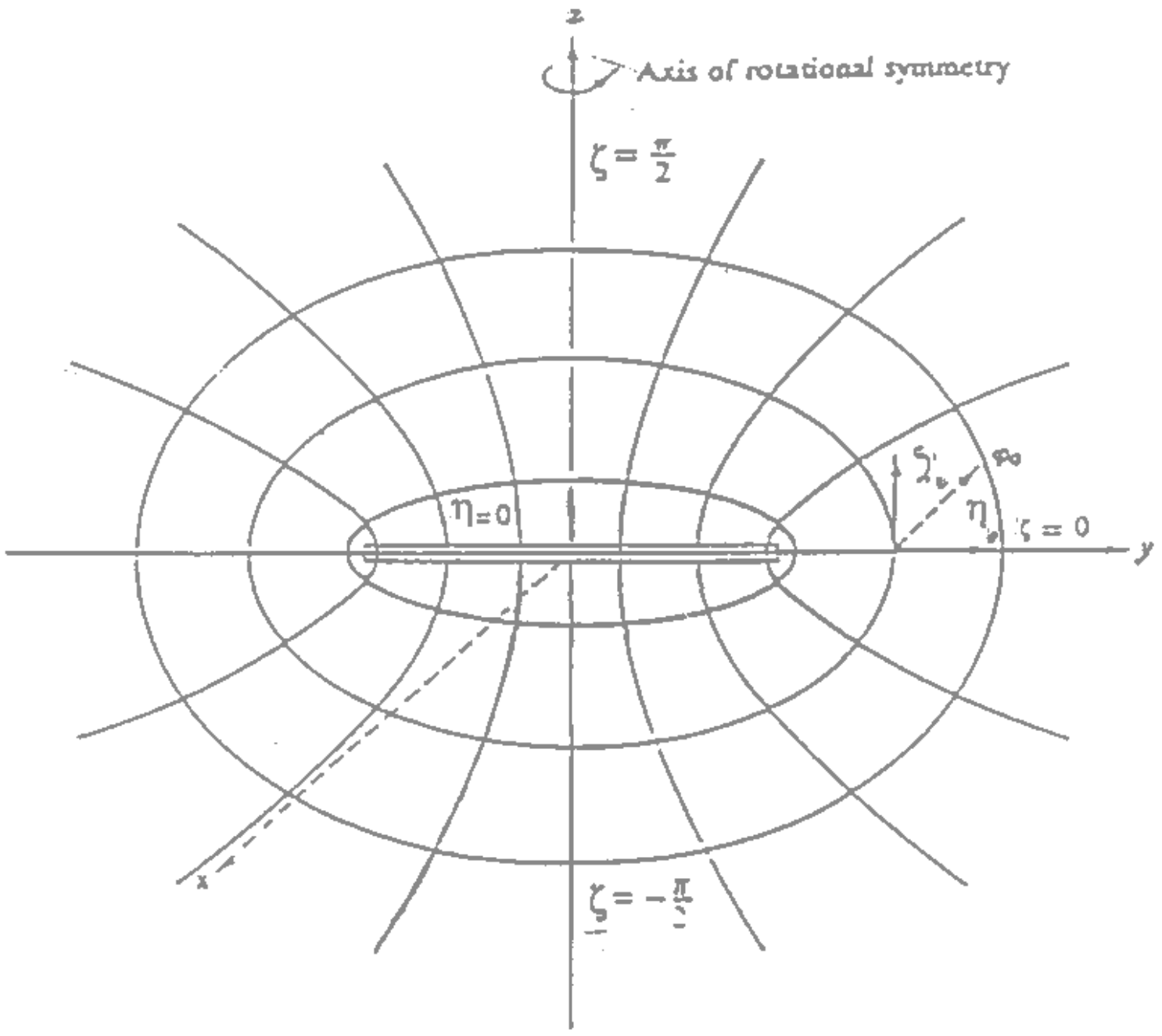

Figure 4. Oblate spheroidal coordinates (from Arfken. [3])

## III.3 Determination of Field Configuration in ElectroPore

We make use of the field conditions that are to be satisfied by the electromagnetic fields:

i.) the normal component of J is continuous across a boundary,

ii.) the tangential component of the electric field is continuous at an interface.

The field can be written in terms of its components as

$$\vec{E}(\eta, \zeta) = E_\eta(\eta, \zeta)\hat{\eta} + E_\zeta(\eta, \zeta)\hat{\zeta}$$

For steady state conditions, the fields are found to be of the forms

$$E_\eta(\eta, \zeta) = \frac{E_0^\eta}{\left(1 + \eta^2\right)^{\frac{1}{2}} \left(\eta^2 + \zeta^2\right)^{\frac{1}{2}}} \quad (3.1)$$

$$E_\zeta(\eta, \zeta) = \frac{E_\zeta^0}{\left(1 - \zeta^2\right)^{\frac{1}{2}} \left(\eta^2 + \zeta^2\right)^{\frac{1}{2}}} \quad (3.2)$$

The z axis corresponds to the value of $\zeta = 1$, and is included in the region of the pore in the membrane. From the expression for $E_\zeta(\eta, \zeta)$, points on the z axis would have an electric field with a magnitude which is undefined, thus we take $E_\zeta^0 = 0$ . We see therefore , that in the pore, the field is entirely in the $\eta$ direction.

## III.4 Curvature of the membrane

In order to find the potential distribution on the membrane we must first find some way to describe the membrane in the coordinate system chosen. In an attempt to do this, the curvature of the edge of the membrane when the pore is formed will be calculated. This curvature for a particular value of the coordinate $\xi$ gives the shape of the membrane in the region of the pore. We will take this shape to be consistent independent of the size of the pore. From the relational equations, the shape of this curve z can be written as a function of $\rho$ viz.

$$Z(\rho) = \frac{\zeta}{\left(1 - \zeta^2\right)^{\frac{1}{2}}} \left( \rho^2 - a^2\left(1 - \zeta^2\right)\right)^{\frac{1}{2}} \quad (3.3)$$

The curvature K of this curve can be found from the definition

$$K = \frac{\dfrac{d^2 Z}{d\rho^2}}{\left(1 + \left(\dfrac{dZ}{d\rho}\right)^2\right)^{\frac{3}{2}}} \qquad (3.4)$$

Writing the apex pore radius as

$$a^2\left(1-\zeta^2\right) = b^2 \qquad (3.5)$$

a direct calculation for K gives

$$K = \frac{-\left(1-\zeta^2\right)\,\zeta\, b^2}{\left[\left(1-\zeta^2\right)\left(\rho^2-b^2\right) + \zeta^2\rho^2\right]^{\frac{3}{2}}} \qquad (3.6)$$

From this expression for K, it can be determined that the maximum curvature will occur at the point $\rho = b$, for which the maximum curvature is found to be,

$$K_m = \frac{-\left(1-\zeta^2\right)}{\zeta^2 b} \qquad (3.7)$$

The membrane must be described by a $\zeta$ that approximates the real physical parameters of the membrane. The real physical parameters of the membrane, including the length of the lipid molecules, the closest separation between the phosphor heads, the separation of the lipid tails and the thickness of the membrane, must be related to the particular value $\zeta_0$ used to describe the membrane. The curves of constant $\zeta$ result in a z value that is a monotonically increasing function, and it is thus necessary to determine the value of the parameter $\eta_o$ for which the thickness of the model membrane is equal to the true membrane thickness, t. This value of $\eta$ forms the limit for which the use of the oblate spheroidal coordinates is applicable. The point denoted by ( $\eta_0$, $\zeta_0$) corresponds to a distance $\rho_0$ from the axis of the pore . With the distance from the axis through the pore given by $\rho$ and parametrizing the position at which the membrane thickness is the required value, it is possible to write $\dfrac{dz}{d\rho}$, $\dfrac{d^2 z}{d\rho^2}$ and K in terms of the parametrization $\rho = b + \lambda$ . Using the region of maximum curvature , $K_m$ we obtain

$$K_m b = \frac{-\left(1-\zeta_0^2\right)}{\zeta_0^2}$$

and in terms of $\rho = b + \lambda$ the terms $\frac{dZ}{d\rho}$ , $\frac{d^2 Z}{d\rho^2}$ and K can be expressed as,

$$\frac{dz}{d\rho} = \left(\frac{1}{K_m b}\right)^{\frac{1}{2}} \frac{(b+\lambda)}{\left(2b\lambda+\lambda^2\right)^{\frac{1}{2}}} \quad (3.8)$$

$$\frac{d^2 z}{d\rho^2} = \left(\frac{1}{K_m b}\right)^{\frac{1}{2}} \left(\frac{-b^2}{\left(2b\lambda + \lambda^2\right)^{\frac{3}{2}}}\right) \quad (3.9)$$

$$K = -\frac{K_m b \left(- b^2\right)}{\left[\left(K_m b\right)\left(2b\lambda + \lambda^2\right) + (b+\lambda)^2\right]^{\frac{3}{2}}} \quad (3.10)$$

In order to determine the particular choice of the parameter $\zeta_0$ that will be used to describe the membrane, we will use certain properties of the membrane components. The phospo-lipid molecules that comprise the membrane have a given length , L. The minimum separation of the phospo heads in the membrane is given by a distance, d.

We can construct an approximation to the membrane shape around the pore by arranging the phosphor-lipid molecules such that the minimum head separation is d, with the closest approach between the tails of adjacent lipids constrained to a minimum distance $\delta$. Using this approach, a model of the membrane can be constructed that resembles the properties of the membrane sufficiently far away from the pore. Figure 7 illustrates this model.

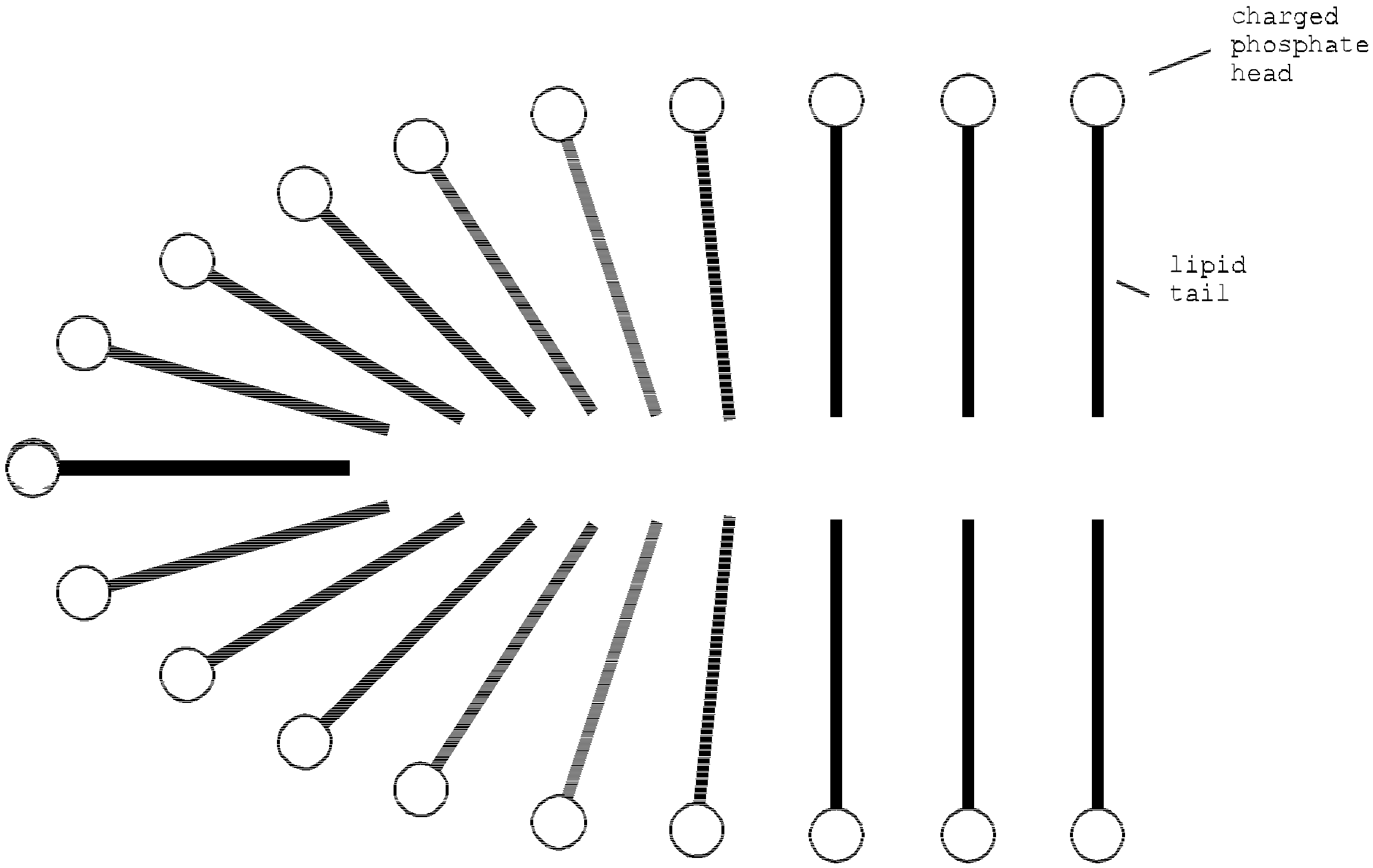

Figure 7. Schematic of Lipid Layout in the Membrane Pore.

The equation of the curve, however cannot be determined from purely geometrical requirements, and we make an approximation for the curve based on the maximum curvature of the membrane determined from the biophysical membrane parameters. We need to find some way of defining the particular ζ that has the properties required of the membrane, primarily that the thickness of the membrane remains relatively constant at a value t. We determine the curvature of the curve based on the model shown in figure 8.

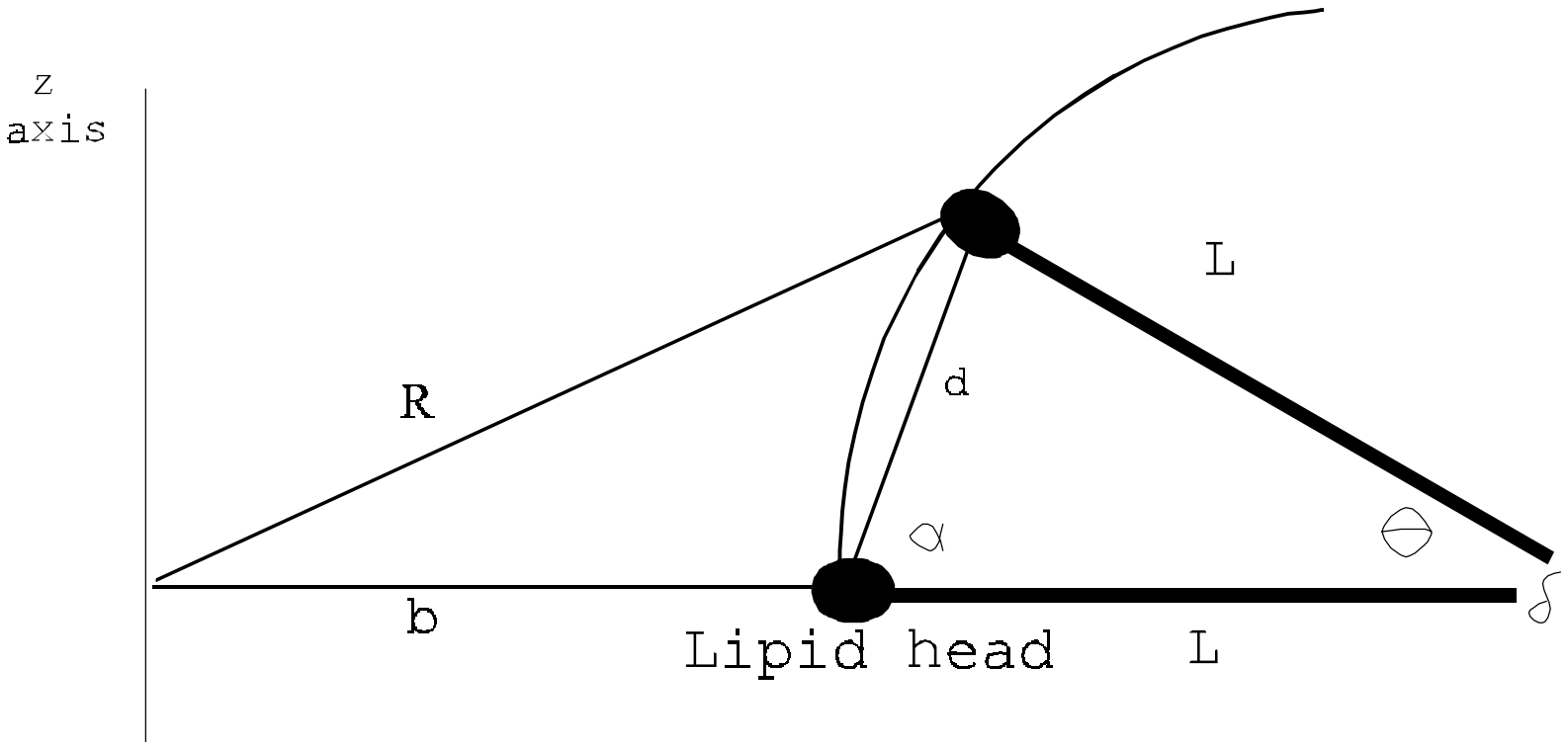

Figure 8. Determination of Membrane Curvature

In this model we assume that we can construct the membrane from the phospho-lipid molecules starting with one horizontal and placing the adjacent molecules so that the heads are a minimum distance d apart, and the tails have a closest distance $\delta$. Based on this model adjacent molecules might not have the same center, however since $\delta$ is ~ 0.1 L, we can make the approximation that the centers are the same. From the figure we can relate the angles by

$$\theta = 180 - 2\alpha$$

Thus we can express the distances from some fixed point, chosen to be the origin, to the head of the adjacent molecule as

$$x = b + L - L\cos\theta \quad ; \quad y = L\sin\theta$$

This means that the position of the head of the adjacent lipid can be expressed as

$$R(\theta) = b + L(1 - \cos\theta)\,\hat{i} + L\sin\theta\,\hat{j}$$

The tangent vector to the curve is given by

$$\vec{T}(\theta) = \frac{dR(\theta)}{d\theta} = L\left[\sin\theta\,\hat{i} + \cos\theta\,\hat{j}\right]$$

From the parametric expression for the curvature

$$K = \frac{\left|\frac{d\hat{T}(\theta)}{d\theta}\right|}{\frac{ds}{d\theta}}$$

which is calculated to be

$$K = \frac{1}{L}$$

From the previous expression for the maximum curvature of the curve $\zeta_0$, we can now calculate the value $\zeta_0$

$$K_m = \frac{\left(1 - \zeta_0^2\right)}{\zeta_0^2\, b} = \frac{1}{L} \qquad (3.11)$$

solving this equation for $\zeta_0$ in terms of L and b gives

$$\zeta_0 = \left( \frac{L}{L + b} \right)^{\frac{1}{2}} \qquad (3.12)$$

We express the membrane thickness t, in terms of the parameters L, δ thus

$$t = 2L + \delta$$

We will henceforth assume that δ is much less than L, and can be ignored.

We can now find the value of $\eta_0$ for which the membrane thickness is attained. When the membrane has a thickness t, we can write the z coordinate of the parametric curve to be given by z=L. Using the transformation equation to oblate spheroidal coordinates

$$z = a \eta \zeta$$

Substitution gives

$$L = a \eta_0 \left( \frac{L}{L + b} \right)^{\frac{1}{2}}$$

Using the relationship

$$b = a\left(1 - \zeta_0^2\right)^{\frac{1}{2}}$$

we obtain the expression for the coordinate parameter a expressed in terms of the biophysical parameters desired

$$a = b^{\frac{1}{2}} \left(L + b\right)^{\frac{1}{2}} \qquad (3.12)$$

Using $\zeta_0$ from 3.12 we can write $\eta_0$ as

$$\eta_0 = \left( \frac{L}{b} \right)^{\frac{1}{2}} \qquad (3.13)$$

We can find the distance from the z axis to the position where the thickness of the membrane reaches the value L. This distance is given by

$$\rho_0 = \left(b \left( b + L\right) \right)^{\frac{1}{2}}$$

At this location, the slope of the membrane is found from equation (3.8) as

$$\frac{dz}{d\rho} = \left(\frac{b + L}{b}\right)^{\frac{1}{2}} \qquad (3.14)$$

which gives a measure of the geometrical mismatch between the oblate spheroidal pore and the membrane which away from the pore has vanishing slope.

At the position $(\eta_0, \zeta_0)$ the geometrical bound of the electropore described in oblate spheroidal coordinates meets the remainder of the membrane described in spherical coordinates, which, for regions close to the pore, may be taken as being essentially flat. From the slope we see that the curve makes a non-negligible angle in the region where it transitions to the remainder of the membrane. Thus at the edge there is a small region of mismatch between the fields, represented in one region by oblate spheroidal coordinates and in the other region by spherical coordinates. We must therefore relate the fields in the two different regions from a consideration of the continuity of the components of the field. The quantity expressed in the two different coordinate systems represents the same field, thus the magnitude and direction of the field in each region should approximately match that in the other region. At the edge $(\eta_0, \zeta_0)$ the fields in the two regions can be written in terms of the its components in Cartesian coordinates.

In the region asymptotic represented by spherical coordinates, the field $\vec{E}$ can be expressed as $\vec{E} = E_r\hat{r} + E_\theta\hat{\theta}$. These components can then be expressed in terms of cartesian coordinates as,

$$E_r \Rightarrow E_r\cos(90-\theta)\hat{x} + E_r\sin(90-\theta)\hat{y}$$

$$E_\theta \Rightarrow E_\theta\cos\theta\,\hat{x} - E_\theta\sin\theta\,\hat{y}$$

The components of the field in spherical coordinates are found from the potential distribution found,

$$E_r = -\frac{\partial V_{II}}{\partial r}$$

$$E_\theta = -\frac{1}{r}\frac{\partial V_{II}}{\partial \theta}$$

with $V_{II}$ given in the reference [36] (Hercules, et.al., 2002). The field is given by

$$E_r = \left[ \frac{3\,\sigma_1 R \left(1 - 2\,\frac{t}{R}\right)}{\left(3R\sigma_2 + 2t\sigma_1\right)} \right] E_0 \cos\theta$$

$$E_\theta = \left[ \frac{-\;3\sigma_1 R \left(\frac{\sigma_2}{\sigma_1} + \frac{t}{R}\right)}{\left(3R\sigma_2 + 2t\sigma_1\right)} \right] E_0 \sin\theta$$

We can then calculate the magnitude of the vector $|E| = \left(E_r^2 + E_\theta^2\right)^{\frac{1}{2}}$ . Ignoring terms of second order in small parameters, this gives the magnitude of the vector in spherical coordinates as

$$\left|E^{sp}\right| \simeq \frac{3\sigma_1 R\left(1 - 4\frac{t}{R}\right)^{\frac{1}{2}}}{\left(3R\sigma_2 + 2t\sigma_1\right)} E_0 \cos\theta$$

with, $R$ equal to the cell hemisphere radius, which is taken to be the same as the radius of the filter pore, λ. In oblate spheroidal coordinates, the field at the position $\left(\eta_0, \zeta_0\right)$ is given by equation (3.1)

$$E_\eta = \frac{E_\eta^0}{\left(1 + \eta_0^2\right)^{\frac{1}{2}}\left(\eta_0^2 + \zeta_0^2\right)^{\frac{1}{2}}}$$

This can be written in terms of the parameters L and b, using the values found for $\eta_0$ ( 3.12) and $\zeta_0$ (3.13). This gives

$$E_\eta = \frac{E_\eta^0\; b}{\left[L\left(L + 2b\right)\right]^{\frac{1}{2}}} \qquad (3.15)$$

Setting the magnitudes of the fields to be equal, we have

$$\frac{3\sigma_1 R\left(1 - 4\frac{t}{R}\right)^{\frac{1}{2}}}{\left(3R\sigma_2 + 2t\sigma_1\right)} E_0 \cos\theta = \frac{E_\eta^0\; b}{\left[L\left(L+2b\right)\right]^{\frac{1}{2}}}$$

where, $R$, in this expression is set equal to, $\lambda$. Thus the field in the near region of the remainder of the membrane is related to the field in the electropore by

$$E_0 = \frac{E_\eta^0\, b}{\left[L(L+2b)\right]^{\frac{1}{2}}}\; \frac{\left(3R\sigma_2 + 2t\sigma_1\right)}{3\sigma_1 R\left(1 - 4\frac{t}{R}\right)^{\frac{1}{2}} \cos\theta} \tag{3.16}$$

At the position $\left(\eta_0, \zeta_0\right)$ the polar angle, denoted $\theta_0$, can be found from the relation

$$\cos\theta_0 = \frac{\left(R^2 - \rho_0^2\right)^{\frac{1}{2}}}{R} = \frac{\left(\lambda^2 - \rho_0^2\right)^{\frac{1}{2}}}{\lambda}$$

Substituting for $\rho_0$ gives

$$\cos\theta_0 = \left(1 - \frac{b(L+b)}{\lambda^2}\right)^{\frac{1}{2}}$$

Thus the field in the region of coordinate matching $E_0$ becomes

$$E_0 = \frac{E_\eta^0\, b}{\left[L(L+2b)\right]^{\frac{1}{2}}}\; \frac{\left(3\lambda\sigma_2 + 2t\sigma_1\right)}{3\sigma_1\left(1 - 2\frac{t}{\lambda}\right)^{\frac{1}{2}}}\; \frac{1}{\left[\lambda^2 - b(L+b)\right]^{\frac{1}{2}}} \tag{3.17}$$

We can use this expression to calculate the current flowing through the remainder of the membrane.

## III.5 Shear forces in the membrane

We determine the relationship between the magnitude of the electric field and the pore size by relating the shear force between adjacent molecules in the lipid bilayer to the force due to the surface tension in the membrane. We then relate the field in the electropore to the applied field in the chamber, and thus we establish the relationship between the applied field and the pore radius.

The force between any two charges in the membrane due to the electric field must be calculated in order to relate the electric field to the size of the pore. The shear force on adjacent charge sites is given by

$$\vec{F}_{sh} = \vec{F}_2 - \vec{F}_1 = \left(\vec{E}_2 - \vec{E}_1\right)q$$

where $E_2$ , and $E_1$ are the effective electric fields at the different positions, $r_1$, $r_2$.
This difference in the fields between the charges can be written as

$$\vec{E}_2 - \vec{E}_1 = \Delta\vec{E} = \vec{\nabla}\vec{E}\,.\delta\vec{r}$$

Thus the shear force in the membrane can be written as

$$\vec{F}_{sh} = \vec{\nabla}_r\vec{E}\,.\delta\vec{r}\; q$$

where $\delta\vec{r} = \vec{r}_2 - \vec{r}_1$. The shear force in the membrane is along the surface of the membrane in the direction of $\eta$. This means that the directional derivative is the gradient in the direction of $\eta$.

$$\vec{\nabla}_\eta = \frac{1}{h_\eta}\frac{\partial}{\partial\eta}$$

(where h is the local scale factor) so that the shear force is

$$F_{sh} = q\cdot d\,\nabla_\eta E_\eta$$

The charge separation d corresponds to the separation of the polar heads of the lipid molecules. This separation is the distance of closest approach of the phospholipid heads. The force opposing the shear effect is due to the surface tension in the membrane. The force due to the surface tension is typically written in the form

$$\vec{F}_{st} = \gamma\, l$$

where l = circumference of the pore and $\gamma$ is defined to be the surface tension of the membrane. In this case,

$$l = 2\,\pi\,\rho$$

where $\rho=\rho_o$ is the previously calculated radius of the pore at the position where the thickness of the membrane has the value *t*. This gives a force from surface tension of

$$F_{st} = 2\,\pi\,\gamma\left(b\,(b+L)\right)^{\frac{1}{2}}$$

From the form of the electric field found previously in equation (3.1) and the transformation factor h we have

$$\nabla_\eta\cdot E_\eta = \frac{1}{h_\eta}\frac{\partial}{\partial\eta}E_\eta = \frac{1}{a}\left(\frac{1+\eta^2}{\eta^2+\zeta^2}\right)^{\frac{1}{2}}\frac{\partial}{\partial\eta}\left(\frac{E_\eta^0}{\left(\eta^2+\zeta^2\right)^{\frac{1}{2}}\left(1+\eta^2\right)^{\frac{1}{2}}}\right)$$

$$= \frac{-E_{\eta}^{0}\ \eta}{a\left(\eta^{2}+\zeta^{2}\right)^{2}}\left(\frac{1+2\eta^{2}+\zeta^{2}}{\left(1+\eta^{2}\right)}\right)$$

Equating the shear force in the membrane to the surface tension yields

$$\frac{-E_{0}^{\eta}\ \eta}{a\left(\eta^{2}+\zeta_{0}^{2}\right)^{2}}\left[\frac{1+2\eta^{2}+\zeta_{0}^{2}}{\left(1+\eta^{2}\right)}\right] q.\ d = 2\ \pi\ \rho\ \gamma \qquad (3.18)$$

This can be written in terms of the parameters b and L as

$$\frac{-E_{0}^{\eta}\ \ L^{\frac{1}{2}}\left[b(L+b)\right]^{2}}{\left[b(L+b)\right]^{\frac{1}{2}}\ b^{\frac{1}{2}}\left[L(L+2b)\right]^{2}}\left[\frac{L(L+2b)+(L+b)^{2}}{(L+b)^{2}}\right] q\ .d = 2\ \pi\ \gamma\left[b(L+b)\right]^{\frac{1}{2}}$$

This equation relates the electric field in the electropore $E_{\eta}^{0}$ to the radius of the electropore and the surface tension of the membrane, $\gamma$. Thus we find that

$$-E_{\eta}^{0} = \frac{2\ \pi\ \gamma}{q.d}\ \frac{(L+b)\left[L(L+2b)\right]^{2}}{(Lb)^{\frac{1}{2}}\left[L\,(L+2b)+(L+b)^{2}\right]} \qquad (3.19)$$

The field in the electropore can then be related to the externally applied field through the continuity condition giving a relationship between the applied field, the size of the electropore formed, and the surface tension of the membrane. Under the effect of the shear force in the membrane the pore will increase in size until the shear force is balanced by the surface tension due to the change in the pore diameter.

## III.6 Continuity condition

Experimentally, RBC's are trapped within the pores of a nonconducting polycarbonate filter. When an electric field is applied to the end electrodes, a potential difference is established across the filter and there is a current flow within the chamber. For the chamber with filter and cells, the only field boundary conditions that can be written involves current conservation and the asymptotic values of the potential. The applied potential will be assumed as +V at $-Z_0$ and -V at $+Z_0$ .

The current flowing through the circuit can be directly measured. Since the filter has a large resistance, we can assume that all of the current flows through the pores in the filter. With the cells embedded In the filter, the current flowing through the filter pores must be the same as that flowing through the cell membrane plus the leakage current. In addition, since the asymptotic value of the electric field is known, and the applied field is essentially uniform, the current flowing can be related to the asymptotic field. This relationship is used to relate the magnitude of the field in the pore of the filter to the magnitude of the externally applied field.

The continuity conditions require the current flowing through the solution in the upper and lower portions of the chamber be equal to the current flowing through the pores of the filter. Thus we have that

$$I_{ch} = \sigma_{sal} E_{app} A_{ch} = N I_f$$

where N is the number of filter pores.

When the cells are embedded in the filter pores, the current through the filter pore is the sum of the current through the cell and the leakage current around the cell. We can write this as

$$I_f = I_p + I_l + I_{mem}$$

where
$I_f$ = current through the filter pore
$I_p$ = current through the electropore
$I_l$ = leakage current around the cell
$I_{mem}$ = current through the remainder of the membrane

For the cell embedded in the filter, the application of an external field leads to the flow of current across the cell membrane. Conservation of charge requires that the current flowing through the lower part of the cell membrane be equal to the current through the electropore plus the current through the remainder of the membrane in the filter pore. This condition on the current flow through the cell may be written as,

$$I_{bot} = \int \vec{J}_{bot} . \hat{n} da = \int \vec{J}^{pore}_{top} . \hat{n} \; da + \int \vec{J}^{m}_{top} . \hat{n} da$$

where the terms on the right indicate the contribution through the electropore and the remainder of the membrane at the top.

The integral of the current density in the pore is over the region of the pore formed in the membrane. The limit of the integral of the current density through the membrane is from the radius of the electropore to the radius λ, of the filter pore (λ ~0.925 μm). To determine the limits of the integrals,, the shape of the membrane within the filter pore is assumed to be in the shape of a hemisphere with a radius equal to the radius of the filter pore.

The cells embedded in the filter do not make a perfect seal with the walls of the filter, thus we make allowances for the leakage current around the cell. The leakage current flows through the thin film of liquid that exists between the cell membrane and the edge of the filter. This current is driven by the electric field that exists at the edge of the filter pore.

We want to calculate the current flowing through the membrane pore, using the form of the electric field found previously. The current flow through the pore is in the direction $\eta$ , so from the component of the electric field in the $\eta$ direction we have

$$I_{pore} = \int_{pore} \vec{J} \cdot d\vec{a}_{pore} = \int_{pore} \sigma \vec{E} \cdot d\vec{a}_{pore}$$

The general area element is

$$da_{\zeta\varphi} = ds_\zeta \, ds_\varphi = h_\zeta \, d\zeta \;\; h_\varphi \, d\varphi$$

with the scale factors given by

$$h_\zeta = a\frac{\left(\eta^2 + \zeta^2\right)^{\frac{1}{2}}}{\left(1 - \zeta^2\right)^{\frac{1}{2}}}$$

and

$$h_\varphi = a\left(1 + \eta^2\right)^{\frac{1}{2}}\left(1 - \zeta^2\right)^{\frac{1}{2}}$$

Substitution into the integral gives the current flowing through the membrane pore as

$$I_{pore} = \sigma\int_{\zeta_0}^{1}\int_{0}^{2\pi} \frac{E_\eta^0 \, a^2 \left(\eta^2 + \zeta^2\right)^{\frac{1}{2}} \left(1 + \eta^2\right)^{\frac{1}{2}} \, d\zeta \, d\varphi}{\left(1 + \eta^2\right)^{\frac{1}{2}} \left(\eta^2 + \zeta 2\right)^{\frac{1}{2}}}$$

$$= 2\pi a^2 \sigma E_\eta^0 \left(1 - \zeta_0\right)$$

The upper limit of $\zeta$ is 1, and we have already determined $\zeta_0$ to be

$$\zeta_0 = \left(\frac{L}{L + b}\right)^{\frac{1}{2}}$$

thus the current through the pore is

$$I_{pore} = 2\pi\, a^2\, \sigma\, E_\eta^0 \left[1 - \left(\frac{L}{L + b}\right)^{\frac{1}{2}}\right]$$

Substituting for a from (3.12) gives

$$I_{pore} = 2\,\pi\, b(L+b)\, \sigma\, E_\eta^0 \left[1 - \left(\frac{L}{L+b}\right)^{\frac{1}{2}}\right] \qquad (3.20)$$

We can next calculate the current flowing through the remainder of the membrane, from the edge of the pore to the walls of the filter. The current flow through the membrane is due to the normal component of the field at the membrane, this component is the radial component of the field at the membrane. We assume that for a small electropore, the potential distribution over the membrane is not significantly altered, thus we use the functional form of the potential already found with the coefficients evaluated on the membrane around the electropore. The outer radius of the portion of the cell in the filter pore is equal to the radius of the filter pore, λ. The current through the remainder of the membrane is thus

$$I_{rem} = 2\,\pi\, \sigma_2 \int_{\theta_0}^{\frac{\pi}{2}} E_r\, \lambda^2 \sin\theta \cos\theta\, d\theta$$

The lower limit of this integral, $\theta_0$, is the angle formed by the electropore. We can determine the angle $\theta_1$ from the relation

$$\rho_0 = \lambda\, \theta_0$$

where $\rho_0$ is the radius of the electropore and λ is the radius of the filter pore. Evaluating the integral yields

$$I_{mem} = \pi\, \sigma_2\, E_f\, \lambda^2 \left[1 - \frac{b(L+b)}{\lambda^2}\right] \qquad (3.21)$$

This can be written in terms of the field in the filter pore. Equation (3.17) relates the field in the electropore to the field in the remainder of the membrane, thus in terms of the field in the electropore the current through the remainder of the membrane is

$$I_{mem} = \pi\, \sigma_{mem} \frac{b\, \lambda \left[ \lambda^2 - b(L+b) \right]^{\frac{1}{2}}}{\left[ L(L+2b) \right]^{\frac{1}{2}}} E_{\eta}^{0} \tag{3.22}$$

We must now relate the current in the filter pore to the current in the chamber, using the conditions already stated. We have calculated the current through the electropore and the remainder of the membrane. Since we do not have an analytic expression for the field in the pore of the filter nor the thickness of layer of liquid between the cell and the wall of the filter, the leakage current must be determined from experimental parameters.

The current conservation conditions give

$$I_{ch} = N \left( I_p + I_{mem} + I_L \right)$$

Substituting for $I_{ch}$, $I_p$, and $I_L$ gives the relationship between the applied field and the field in the electropore and the remainder of the membrane

$$E_{\eta}^{0} = \frac{\sigma_3 E_{app} A_{ch} - N I_L}{N \left( \tilde{I}_p + \tilde{I}_{mem} \right)}$$

where $\tilde{I}_p$ and $\tilde{I}_{mem}$ are related to $I_p$ and $I_{mem}$ by the expressions

$$I_p = E_{\eta}^{0} \tilde{I}_p$$

$$I_{mem} = E_{\eta}^{0} \tilde{I}_{mem}$$

We can then relate the applied field to the radius of the electropore formed, b.

Combining the relation between the field in the electropore and the radius of the pore gives

$$E_{app} = \left(\frac{1}{\sigma_{sal} A_{ch}}\right) \left( \frac{2\pi \gamma \left[L(L+2b)\right]^2 (L+b)}{q\ d\ (Lb)^{\frac{1}{2}} \left[L(L+2b) + (L+b)^2\right]} \right) x$$

$$N \left( 2\pi\ \sigma\ b(L+b) \left[ 1 - \left(\frac{L}{L+b}\right)^{\frac{1}{2}} \right] + \frac{\pi\ \sigma_{mem} b\lambda \left[\lambda^2 - b(L+b)\right]^{\frac{1}{2}}}{\left[L(L+2b)\right]^{\frac{1}{2}}} \right)$$

$$+ \frac{N}{\sigma_{sal} A_{ch}} I_{leak} \qquad (3.23a)$$

Substituting the dependance of the leak current on the applied field, we obtain the relation

$$E_{app} = \left(\frac{1}{\sigma_{sal} A_{ch}}\right) \left( \frac{2\pi \gamma \left[L(L+2b)\right]^2 (L+b)}{q\ d\ (Lb)^{\frac{1}{2}} \left[L(L+2b) + (L+b)^2\right]} \right) x$$

$$\frac{N \left( 2\pi\ \sigma\ b(L+b) \left[ 1 - \left(\frac{L}{L+b}\right)^{\frac{1}{2}} \right] + \frac{\pi\ \sigma_{mem} b\lambda \left[\lambda^2 - b(L+b)\right]^{\frac{1}{2}}}{\left[L(L+2b)\right]^{\frac{1}{2}}} \right)}{1 - \frac{N}{\sigma_{sal} A_{ch}} \frac{\text{filter thickness}}{R_{leak}}} \qquad (3.23b)$$

The field required to produce an electropore of a given size may then be found from this relation. We can utilize values found in the literature to obtain a graph of the expected pore size resulting from a given field applied to the chamber. The field applied to the chamber is expected to be multiplied by the factor

$$\text{Field Multiplication Factor} = \frac{1 - \left( \frac{R_{chamber+saline}}{R_{chamber+saline+filter+cells}} \right)}{\text{filter thickness}} \qquad (3.24)$$

to obtain the field that is applied across the filter given in equation (3.23). The surface tension is expected to vary considerably with the local temperature of the membrane. Rand, [63] reported a value of the surface tension of 0.007 - 0.018 dynes/cm. The erythrocyte membrane thickness is approximately 8 nm . With values ranging from about 5 - 10 nm reported. The resistivity of the saline solution is approximately 65 Ω/cm [72] and the resistivity of the saline is approximately 3.5 times that of the cytoplasm [72]. Using these values, the expected relationship between pore size and field applied across the chamber is shown in the graphs of figure 9.

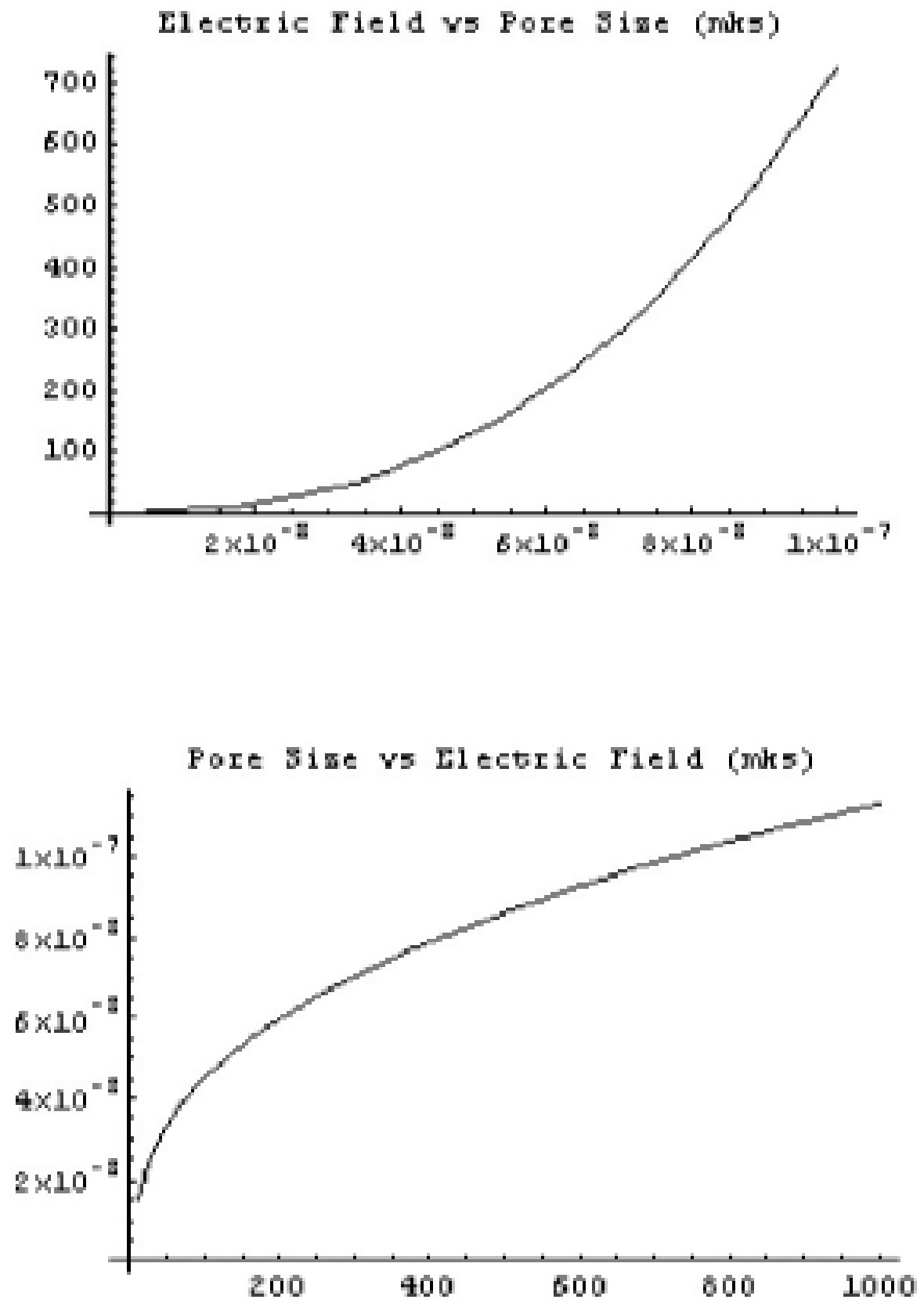

Figures 9a & b. Graphs of pore radius resulting from field applied to chamber

## III.7 Comparison of results with experimental data

Calculation of the pore size from experimental data [72] is shown in the appendix, and gives an approximate pore radius of from $1.69 \le r \le 3.17$ *nm* when using a potential difference of 15V applied to the ends of the chamber. This potential difference generates an electric field of 750 V/m in the chamber. A calculation of the field required to produce a pore of radius 7.5 nm, shows that the field required is approximately 3-6 *V/m* for an ideally operating chamber.

**Chapter IV Discussion and Conclusion**

We can compare the prediction of the theory proposed by Crowley [22] and later modified by Maldarelli and Stebe [49] with that of the present theory. Crowley considered the instability produced in the membrane due to the compression effects of the electric field existing across the membrane. This compression produces an instability in the membrane. For a small steady pressure applied to the ends of an elastic medium, the medium compresses according to Hookes law

$$E\,\frac{dL}{L} = -dP$$

The total compression resulting for a finite pressure p, is obtained by integrating Hooke’s law under the assumption that E is constant. Thus we have

$$E\int_{L}^{l}\frac{dl}{l} = E\ln\left(\frac{l}{L}\right) = -\,p$$

Viewing the membrane as capacitor plates with a spring separating them, the elastic force is balanced by the electric compressive tension, giving

$$-\frac{\epsilon\,V^2}{2\,l^2} = E\ln\left(\frac{l}{L}\right)$$

In this approach it is assumed that the instability develops as a monotonic growth of some small initial disturbance, as opposed to an oscillation of increasing amplitude. To determine the instability of the system, a small disturbance of the system is assumed, and the resulting changes in the electric and elastic stresses are calculated. If the net effect of the change in stress is to accelerate the departure from equilibrium, the system is unstable. We see that when the field acts on the capacitor membrane, it produces a small displacement from equilibrium. If the displacement is compressive, the elastic force will increase, thus tending to return the plates to its original position. As the plates approach each other, however, the electric pressure increases, according to its definition

$$-p_{elec} = \frac{\epsilon\,V^2}{2\,l^2}$$

This increase in the electric pressure tends to pull the plates even closer together. If the electric pressure dominates, the equilibrium is unstable.
At some critical measure of the instability he was able to calculate the critical potential existing across the membrane when the mechanical instability reached its critical value, which he considered the stage at which pores were likely to be formed. This value was found to be given by the expression

$$\frac{\epsilon V_{cr}^2}{2E L^2} \approx 0.18$$

where
$\epsilon$ is the permitivitty of the membrane
$V_{cr}^2$ is the critical transmembrane potential at breakdown
$E$ is the Youngs modulus of the membrane
$L$ is the thickness of the membrane.
The value 0.18 corresponds to the measure of the critical membrane instability.
While this theory may predict the formation of pores in the membrane, it does not provide a relationship between the applied field and the size of the pore formed.

Maldarelli and Stebe, using a modification of Crowley's theory to include the anisotropy of the membrane response to the electrocompressive effects of the electric field derived the relationship giving the critical membrane voltage.

The present theory is similar to that presented by Crowley in that it relates the electric field effects to the mechanical properties of the membrane. In this model, the shear effect of the field is related to the surface tension of the membrane. It might be expected that the surface tension of the membrane would be less than than the compressibility of the membrane, since, to compress the membrane requires a physical overlap of the lipid tails of the membrane components. The shearing of the membrane, however, requires the separation of the molecules in the membrane. The forces between the molecules are mainly Van der Waals forces and hydrogenic bonding. Thus we might expect that the separation of the molecules in the membrane would be more easily accomplished than the compression of the molecules.

Comparison between the theoretical value of the field predicted for a detectable pore size, compared to the experimental value of the field gives essentially no correlation. We can therefore infer that the disruption of the membrane based on a simple mechanical model of the forces in the membrane due to the electrical field does not provide an adequate description of the electroporation phenomenon.

While the model does not take in to account the effects of other structures in the membrane (specifically, proteins) the presence of these structures is unlikely to produce a correction of the orders of magnitude necessary to provide agreement with experiment. The mechanical properties of the membrane may also be greatly affected by these structures. The surface properties of the membrane are affected by the presence of the membrane skeleton, which forms a network of filaments attached to the surface of the cytoplasm and has the function of maintaining the membrane integrity in its passage through the vessels. This membrane skeleton serves to reinforce the lipid bilayer, and allows the membrane to withstand the stresses encountered in passage through the blood vessels while still permitting it considerable flexibility [Warren '89].

While the measured properties of the membrane, based on whole cell measurements [63,64 ], includes the effects of the membrane skeleton, the regions of electroporation may be those parts in between the meshwork of protein fibers. Thus the whole cell measurements may not reflect the properties of the membrane affected by the applied field.

The system in which cells are embedded in a polycarbonate filter provides a unique method for studying the properties of biological cells. The system provides advantages over the usual method of performing electroporation in that the normal high voltages are avoided. The method also allows the use of normal, isotonic saline solution so that the cells are in a natural environment for the study. This method also has the advantage that the cells are held in a fixed position in the applied field.

Due to the difficulties of modeling the membrane as a cylindrical pore in a flat sheet and adequately describing the membrane in cylindrical coordinates, we have chosen to model the membrane using oblate spheroidal coordinates . The fields have been developed, based on the requirements of the field conditions, namely that the field must satisfy Maxwells equations and continuity equation.

A simple view of the membrane has been developed based on the physical parameters of the membrane components , which are primarily phospholipid molecules. The shape of the membrane around the pore formed was determined based on the closest packing of these phospholipid molecules. This shape was determined by considering the curvature of the membrane around the lumen of the pore, the maximum curvature being limited by the minimum separation of the heads and tails of the phospholipid molecules. No attempt has been made here to take into account the presence of other molecules embedded in the membrane such as proteins or carbohydrates.

We determined the effect of the applied electric field on the charged polar heads of the phospholipid molecules in the membrane and showed that the applied field creates a shear force in the membrane which leads to an enlargement of the electropore formed. This effect is opposed by the

surface
tension of the membrane, thus establishing a natural balance between the applied field and the restoring effect of the membrane. A relationship has been derived between the magnitude of the applied field and the size of the electropore formed, given by equation (3.23).

The effects are related to simple parameters of the field and membrane. This model does not attempt to describe the vast body of experimental data that exists, but instead seeks to establish a direct relationship between the field and the pore size formed.

**DEDICATION**

I dedicate this effort to the memory of my Mother and Father, who started me on the journey but did not make it to the end with me. Their ideas of perseverance kept me going through the many difficult times.

**ACKNOWLEDGEMENTS**

I wish to acknowledge the support of the many individuals who made it possible for me to successfully complete this work.

My deepest thanks goes to Dr. Anna Coble for all the support over the many years as well as her advice on this project. I am indebted to my thesis advisor Dr. James Lindesay for the guidance provided along the way. Many thanks to Dr. Robert Schmukler for allowing me the use of his laboratory and equipment as well as the many valuable suggestions made. Thanks also to Dr. Mark Esrick for the time and effort he provided on this project. Much thanks to Dr. Tristan Hübsch for his continued encouragement and assistance, as well as the facilities so willingly made available for my use, and his willingness to listen to my ideas and make suggestions. Special thanks to Dr. Demetrius Venable for providing support and encouragement. To Dr. Lewis Klein for unhesitatingly making his resources available to me. To Dr. Robert Catchings for providing necessary support.

I am indebted to the Department of Physics and Astronomy and the Graduate School of Arts and Sciences for their continued support.

I also extend my gratitude to the staff of the Department of Physics and Astronomy especially Ms. Mary-ann Tann for the many problems which she helped solve, and Mr. Ronald Talley for making my chamber.

I am grateful to Dr. Walter Lowe for making his facilities and staff at MHATT-CAT available to assist, in particular I am deeply grateful to Mr. Steven Bassen for the time and care put into making my chambers.

Special thanks go to Dr. Winston Anderson for his willingness in providing materials for the experiment.

Special thanks to my brothers, Merrydale and Robert, and to my sister Margaret, for their considerable support over the years.

Finally, I am eternally indebted to Lisa Charles for her unending love and unconditional support over the duration of this undertaking, without her support this work would never have been completed.

**Appendix**

Calculation of the approximate pore size from experimental data

$Z_0$ = Impedance of chamber with saline ≃ 24 Ω

$Z_1$ = Impedance of chamber with saline and filter = 42 Ω

$Z_2$ = Impedance of chamber with filter and cells before poration = 192 Ω

$Z_3$ = Impedance of chamber with filter and cells after poration = 180 Ω

The leakage impedance $Z_L$ is determined from

$$Z_L = Z_2 - Z_0$$

$Z_L$ = 192 - 24 = 168 Ω

The leakage impedance is due to N parallel leakage resistances. The leakage impedance is in series with the filter and saline impedance. We can therefore say that

$$\frac{1}{Z_L} = \frac{N}{R_L}$$

so

$$R_L = N\,Z_L = 3.3 \times 10^5 \,.\, 168 = 55.4 \times 10^6\ \Omega$$

This after poration the conductance of the membrane changes due to the formation of the pore. The total impedance of the system is now due to the impedance of the saline, plus the impedance of the filter now with the parallel combination of the leak resistance and the impedance of the pore in the cell membrane. We can thus write the total impedance after poration as

$$Z_3 = Z_1 + (\ Z_L // Z_P\ )$$

where $(\ Z_L // Z_p\ )$ = $Z_c$ indicates the parallel combination of $Z_L$ and $Z_p$. We can calculate $Z_c$ from the expression

$$Z_3 = Z_0 + Z_c$$

gives

$$Z_c = Z_3 - Z_0 = 180 - 24 = 156\ \Omega$$

Writing $Z_c$ in terms of $Z_L$ and $Z_p$ we have

$$\frac{1}{Z_c} = \frac{1}{Z_L} + \frac{1}{Z_p}$$

from which

$$Z_p = \frac{Z_L Z_c}{Z_L - Z_c} = \frac{168 \times 156}{168 - 156} = 2.184 \times 10^3\ \Omega$$

The total impedance of the pores is due to N pores in parallel, thus the resistance

of each pore is

$$R_p = N\,Z_p = 3.3 \times 10^5 \text{ x } 2.184 \times 10^3 = 7.21 \times 10^8 \Omega$$

We can thus estimate the radius of the electropore from a 15 V pulse from the relation

$$R = \rho \frac{L}{A}$$

where

$\rho$ = resistivity of the saline =65 $\Omega$ cm [73]

$L$ = thickness of the membrane = t $\approx 10 \times 10^{-9}\, m$

$A$ = cross sectional area of the electropore

and $R_p = 7.21 \text{ x} 10^8\ \Omega$

This gives the radius of the pore as
r =1.69 nm

We can find the range of pore size if we consider that the electropore may be filled with intracellular fluid, which has a resistivity approximately 3.5 times that of the saline. Using this value of the resistivity

$\rho$ = 227.5 $\Omega$ cm

gives the size of the pore as
r = 3.17 nm

When an 8 V pulse is used, experimentally it is difficult to detect the formation of an electropore using the changes in the impedance of the system. With an instrument sensitivity of ~$10^3$ we can estimate the size of the pore that would be electrically undetectable. Since the electropore is in parallel with the leak resistance we expect that a pore resistance greater than approximately $10^3\ R_L$ would be undetectable. This gives an upper limit of the pore resistance of detectable pores of

$R_p' = 10^3\ R_L = 10^3 \text{ x } 5.54 \text{ x } 10^7\ \Omega$ which gives $R_p' = 5.54 \text{ x } 10^{10}\ \Omega$.

We can use this to determine the limits of the radius of a pore that can be electrically detected. For the pore filled with solution of resistivity 65 $\Omega$ cm we find

$R_p' = 5.54 \text{ x } 10^{10}\ \Omega$

$$r^2 = \frac{\rho\, L}{\pi\, R_p'} = \frac{65 \text{ x } 10 \text{ x} 10^{-7}}{\pi \text{ x } 5.54 \text{ x} 10^{10}} = 3.73 \text{ x } 10^{-16}$$

for which we find

$r = 1.9 \text{ x } 10^{-8}\ cm = 0.19\ nm$

With a solution resistivity of 227.5 $\Omega$ cm we get

for which we find

$$r^2 = \frac{\rho L}{\pi R_p^{'}} = \frac{227.5 \, x10 \, x10^{-7}}{\pi \, x \, 5.54x10^{10}} = 1.307 \, x \, 10^{-15}$$

$r = 3.6x10^{-8}$ *cm* = 0.36 *nm*

The radius of the pore is therefore on the order of the size of separation of the lipids in the membrane or the water channels in the membrane. We can therefore expect that these pores would be undetectable from the other similar structures.

We find that at an applied potential of 15 V, the electropore formed has limits of $1.69 \leq r_{15} \leq 3.17$ *nm*.

At an applied potential of 8V we find that the pores are undetectable which gives pore limits of $0.19 \leq r_8 \leq 0.36$ *nm*.